\pdfoutput=1
\documentclass[aps,print,superscriptaddress]{revtex4-1}

\usepackage[separate-uncertainty=true]{siunitx}
\usepackage{times}
\usepackage{graphicx}
\usepackage{bmpsize}
\usepackage{lineno}
\usepackage{float}
\usepackage{amssymb}
\usepackage{mathrsfs}
\usepackage{amsmath, amssymb,textcomp,calc,capt-of,ifthen}
\usepackage[english]{babel}
\usepackage{xcolor}
\usepackage{ulem}
\usepackage{lineno}

\begin{document}

\title{Remote entanglement via adiabatic passage using a tunably-dissipative quantum communication system}

\date{May 27, 2020}

\author{H.-S. Chang}
\affiliation{Pritzker School of Molecular Engineering, University of Chicago, Chicago IL 60637, USA}
\author{Y. P. Zhong}
\affiliation{Pritzker School of Molecular Engineering, University of Chicago, Chicago IL 60637, USA}
\author{A. Bienfait}
\altaffiliation[Present address: ]{Universit\'e de Lyon, ENS de Lyon, Universit\'e Claude Bernard, CNRS, Laboratoire de Physique, F-69342 Lyon, France}
\affiliation{Pritzker School of Molecular Engineering, University of Chicago, Chicago IL 60637, USA}
\author{M.-H. Chou}
\affiliation{Pritzker School of Molecular Engineering, University of Chicago, Chicago IL 60637, USA}
\affiliation{Department of Physics, University of Chicago, Chicago IL 60637, USA}
\author{C. R. Conner}
\affiliation{Pritzker School of Molecular Engineering, University of Chicago, Chicago IL 60637, USA}
\author{\'E. Dumur}
\altaffiliation[Present address: ]{Universit\'e Grenoble Alpes, CEA, INAC-Pheliqs, 38000 Grenoble, France}
\affiliation{Pritzker School of Molecular Engineering, University of Chicago, Chicago IL 60637, USA}
\affiliation{Argonne National Laboratory, Argonne IL 60439, USA}
\author{J. Grebel}
\affiliation{Pritzker School of Molecular Engineering, University of Chicago, Chicago IL 60637, USA}
\author{G. A. Peairs}
\affiliation{Department of Physics, University of California, Santa Barbara CA 93106, USA}
\affiliation{Pritzker School of Molecular Engineering, University of Chicago, Chicago IL 60637, USA}
\author{R. G. Povey}
\affiliation{Pritzker School of Molecular Engineering, University of Chicago, Chicago IL 60637, USA}
\affiliation{Department of Physics, University of Chicago, Chicago IL 60637, USA}
\author{K. J. Satzinger}
\altaffiliation[Present address: ]{Google, Santa Barbara CA 93117, USA.}
\affiliation{Department of Physics, University of California, Santa Barbara CA 93106, USA}
\affiliation{Pritzker School of Molecular Engineering, University of Chicago, Chicago IL 60637, USA}
\author{A. N. Cleland}
\affiliation{Pritzker School of Molecular Engineering, University of Chicago, Chicago IL 60637, USA}
\affiliation{Argonne National Laboratory, Argonne IL 60439, USA}

\begin{abstract}
Effective quantum communication between remote quantum nodes requires high fidelity quantum state transfer and remote entanglement generation. Recent experiments have demonstrated that microwave photons, as well as phonons, can be used to couple superconducting qubits, with a fidelity limited primarily by loss in the communication channel \cite{Kurpiers2018,Axline2018,Campagne2018,Leung2019,Zhong2019,Bienfait2019}. Adiabatic protocols can overcome channel loss by transferring quantum states without populating the lossy communication channel. Here we present a unique superconducting quantum communication system, comprising two superconducting qubits connected by a 0.73 m-long communication channel. Significantly, we can introduce large tunable loss to the channel, allowing exploration of different entanglement protocols in the presence of dissipation. When set for minimum loss in the channel, we demonstrate an adiabatic quantum state transfer protocol that achieves 99\% transfer efficiency as well as the deterministic generation of entangled Bell states with a fidelity of 96\%, all without populating the intervening communication channel, and competitive with a qubit-resonant mode-qubit relay method. We also explore the performance of the adiabatic protocol in the presence of significant channel loss, and show that the adiabatic protocol protects against loss in the channel, achieving higher state transfer and entanglement fidelities than the relay method.
\end{abstract}

\maketitle

%%%%% INTRODUCTION
Remote entanglement of superconducting qubits has recently been demonstrated using both microwave photon- and phonon-mediated communication \cite{Kurpiers2018,Axline2018,Campagne2018,Leung2019,Zhong2019,Bienfait2019}. Many of these demonstrations are limited by loss in the communication channel, due to loss in the various microwave components or intrinsic to the channel itself \cite{Kurpiers2018,Leung2019,Bienfait2019}; similar limitations apply to e.g. optically-based quantum communication systems. Adiabatic protocols analogous to stimulated Raman adiabatic passage (STIRAP) \cite{Vitanov2017,Bergmann2019} can mitigate such loss by adiabatically evolving an eigenstate of the system, using states that are ``dark'' with respect to the communication channel. These enable the high-fidelity coherent transfer of quantum states between sender and receiver nodes, even in the presence of large channel loss. Despite their use in a number of localized systems, such protocols have not been used for the generation of remote entangled states \cite{Vitanov2017,Bergmann2019}.

In this Letter, we present a unique experimental system comprising a pair of superconducting transmon-style qubits linked by an on-chip, 0.73 m-long superconducting microwave transmission line. By changing the coupling of the transmission line to a resistive load, we can vary the transmission line's energy lifetime $T_{1r}$ over two orders of magnitude. We demonstrate an adiabatic protocol for quantum communication between the qubit nodes, compare its performance to a qubit-transmission mode-qubit relay method \cite{Sillanpaa2007,Ansmann2009,Zhong2019}, and explore the performance of both protocols as a function of transmission loss.

We first describe the experimental device, then the two state transfer methods. We test the performance of each protocol in the low-loss limit, then as a function of transmission loss. The adiabatic process achieves significantly improved performance compared to the relay method, especially at intermediate levels of loss in the channel.

\begin{figure}[t]
	\centering
	\includegraphics[width=8.6cm]{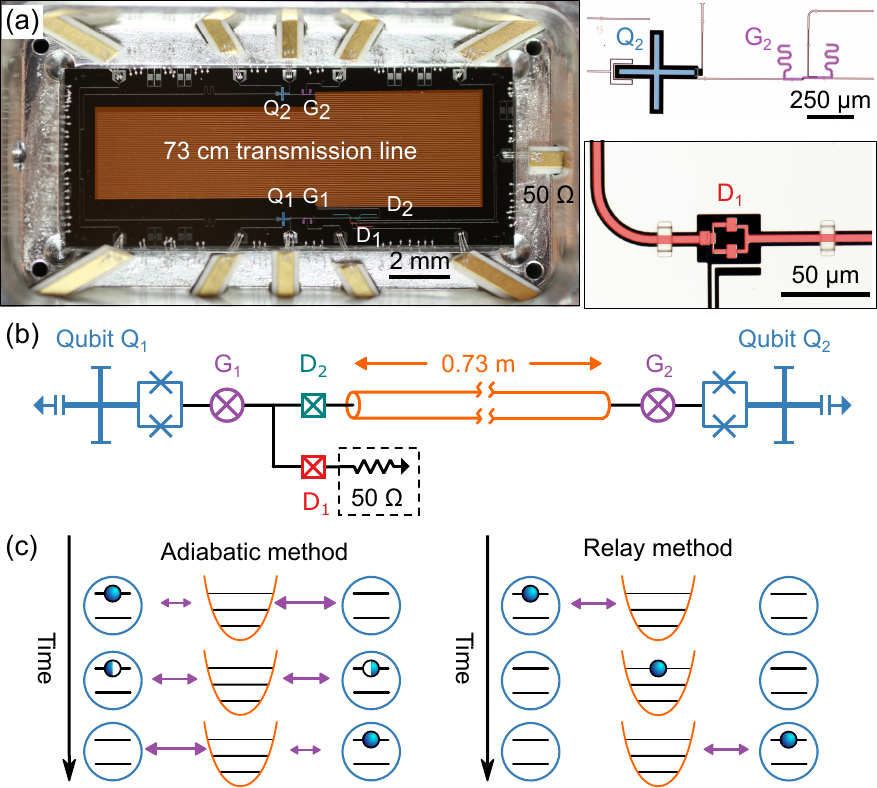}
	\caption{
		\label{fig:device}
		Experimental device.
		(a) Optical micrograph of the device (left), with magnified views of one qubit and its associated tunable coupler (right top), and one variable loss coupler (right bottom).
        (b) A simplified circuit schematic, with two superconducting qubits ($Q_1$ and $Q_2$, blue), coupled by tunable couplers ($G_1$ and $G_2$, purple) to a 0.73 m-long superconducting transmission line (orange). The transmission line is interrupted near $Q_1$ by a tunable switch. The switch comprises two tunable couplers $D_1$ (red) and $D_2$ (teal), with $D_1$ connected to an external $50~\Omega$ load to ground (dashed box), while $D_2$ connects to the remainder of the transmission line. Complete circuit diagram and parameters are provided in \cite{SupplementaryMaterial}.
    }
    \centering
\end{figure}

%%%%% FIG 1 PART
The two quantum state transfer methods, and the device we use to test them, are shown in Fig.~\ref{fig:device}. The device comprises two frequency-tunable superconducting xmon qubits \cite{Koch2007,Barends2013}, $Q_1$ and $Q_2$, each coupled to one end of the on-chip transmission line via an electrically-controlled tunable coupler \cite{Chen2014}, $G_1$ and $G_2$ respectively (Fig.~\ref{fig:device}b). We use the qubit ground $|g\rangle$ and excited $|e\rangle$ states, whose transition frequency is tunable from $\sim$3 to 6 GHz. Qubit control is via low-frequency flux-tuning for $Z$ control and quadrature-resolved microwave pulses for $XY$ control. We read out the qubit states using standard dispersive measurements \cite{Schuster2005,Wallraff2005,Blais2004}, via a capacitively-coupled readout resonator and a traveling-wave parametric amplifier. We projectively measure the excited state probability $P_e$ of each qubit with a fidelity of 88.8$\pm$0.8\%.

The tunable couplers $G_1$ and $G_2$ allow us to externally control the coupling $g_{1,2}$ of each qubit to the individual resonant modes in the transmission line. A variable control consisting of two additional tunable couplers, $D_1$ and $D_2$, is integrated into the transmission line, 1.6 mm from the coupler $G_1$ and its associated qubit $Q_1$. This circuit element provides electrically-controlled coupling between its input port and two output ports \cite{Chang2020}. The coupler $D_2$ is placed inline with the transmission line and is always set to provide maximum coupling (and minimal reflection) to the remaining length of transmission line. The other coupler $D_1$ connects to port 1 on the sample mount, which is terminated by a lumped $50~\Omega$ microwave load outside the sample box. Varying the coupling to this load allows us to set the loss in the transmission line, quantified by the energy lifetime $T_{1r}$ of each resonant mode.

\begin{figure}[t]
	\centering
	\includegraphics[width=8.6cm]{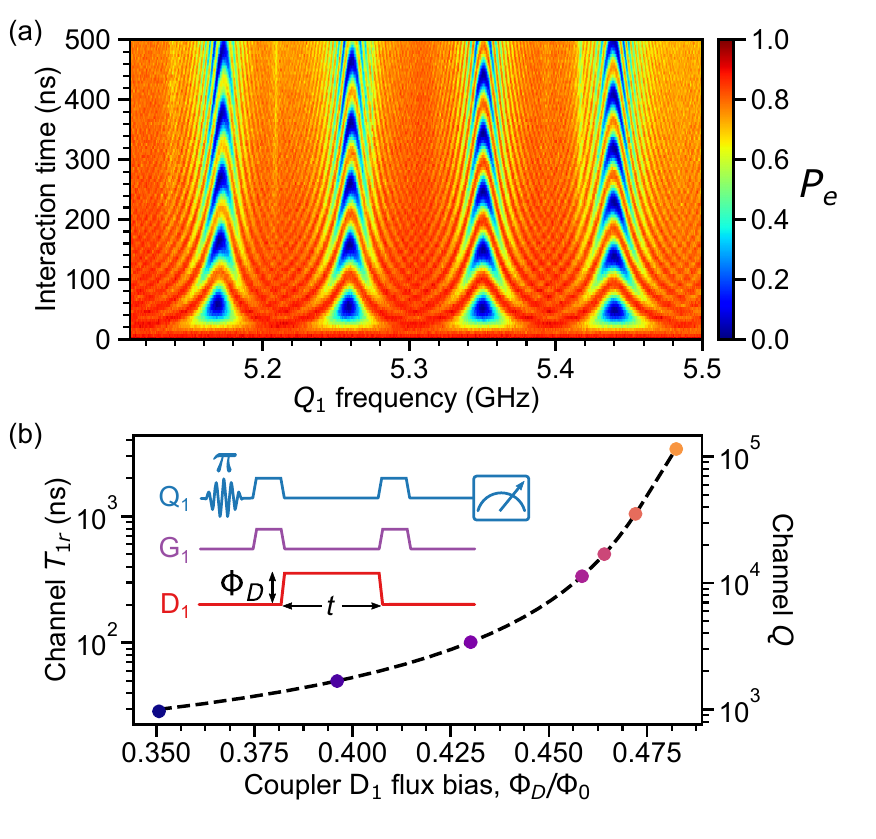}
	\caption{
		\label{fig:channel}
		Variable loss transmission channel.
		(a) Vacuum Rabi swaps between qubit $Q_1$ and four sequential resonant transmission line modes. The coupling is set to $|g_1|/2\pi =5.0\pm0.1$~MHz  $\ll \omega_{\mathrm{FSR}}/2\pi$.
	(b) Measurement of the energy lifetime $T_{1r}$ of one resonant mode in the transmission line, at 5.351 GHz, with equivalent quality factors $Q$ shown on right; inset shows pulse sequence. A $\pi$ pulse to qubit $Q_1$ puts it in the excited state, and this excitation is swapped into the resonant mode for a time $t$, after which it is recovered and the qubit $P_e$ measured. The corresponding lifetime is measured as a function of transmission line loss, controlled during the lifetime measurement using coupler $D_1$. With $D_1$ turned off, we find the intrinsic lifetime $T_{1r}=3410\pm40$~ns (orange); with maximum loss, we find $T_{1r}= 28.7\pm 0.2$~ns (blue). The standard deviation of each data point is smaller than the points. Dashed lines are results calculated with a circuit model; see \cite{SupplementaryMaterial}.
	}
	\centering
\end{figure}

%%%%% FIG 2 PART
The transmission line of length $\ell=0.73$~m supports multiple resonant modes, separated in frequency by the free spectral range $\omega_{\mathrm{FSR}}/2\pi= 1/2T_{\ell} = 84$~MHz, where $T_{\ell} = 5.9$~ns is the photon one-way transit time in the channel. For sufficiently small qubit-resonator coupling, $g_{1,2} \ll \omega_{\mathrm{FSR}}$, each qubit can be selectively coupled to a single resonant mode in the transmission line. This is shown in Fig.~\ref{fig:channel}a, where the transition frequency $\omega_{ge}/2\pi$ of qubit $Q_1$ is tuned over 400 MHz, yielding four separate vacuum Rabi swap resonances spaced by the free spectral range $\omega_{\mathrm{FSR}}/2\pi$. The loss coupler $D_1$ was set to minimum coupling, so the transmission line is limited only by its intrinsic loss. All experiments here were done with the mode at 5.351 GHz, just to the right of center in Fig.~\ref{fig:channel}a.

In Fig.~\ref{fig:channel}b, we demonstrate tunable control over the channel loss, using qubit $Q_1$ to measure the lifetime of the resonant mode at 5.531 GHz as we vary the coupler $D_1$ and thus the transmission line loss. The pulse sequence for this measurement is shown inset in Fig.~\ref{fig:channel}b. The mode energy decay time $T_{1r}$ for each loss setting (controlled by the $D_1$ flux) is shown in Fig.~\ref{fig:channel}b. With no coupling through $D_1$,  we measure the intrinsic resonant mode lifetime $T_{1r} \approx 3410 \pm 40$~ns (orange), comparable to similar transmission lines without variable loss \cite{Zhong2019}. With maximum coupling to the load, we measure a lifetime $T_{1r} \approx  28.7 \pm 0.2$~ns (blue), corresponding to a loaded quality factor $Q_{\mathrm{r}}=960$, about 120 times smaller than the intrinsic quality factor of $1.1 \times 10^5$.  We also measure the resonant mode's Ramsey dephasing time $T_{2r}$ at various $D_1$ flux bias points, and find $T_{2r} \approx 2T_{1r}$, indicating the coupler $D_1$ introduces negligible additional phase decoherence. One non-ideality with this system is that qubit $Q_1$, due to its close proximity to the loss coupler $D_1$, also has its lifetime reduced when the couplers $G_1$ and $D_1$ are both set to non-zero coupling, allowing energy loss from $Q_1$ to the external load; this limits $Q_1$'s performance, and is discussed further in the Supplementary Information \cite{SupplementaryMaterial}. We note that this non-ideality can be avoided by placing the loss coupler $D_1$ in the middle of transmission line, as the transmission line would protect both qubits from the external load.

%%%%% FIG 3 PART
We used two different communication protocols, adiabatic transfer and a qubit-resonant mode-qubit relay method. Both methods were used for qubit state transfer via the transmission line as well as Bell state generation, both as a function of loss in the communication channel. The relay method uses a single extended mode in the transmission line, swapping an excitation from one qubit into that mode and subsequently swapping the excitation from that mode to the other qubit. This method is described in detail elsewhere \cite{Zhong2019}; here it achieves an intrinsic loss-limited state transfer efficiency of $\eta = 0.95 \pm 0.01$ and a Bell state fidelity of $\mathcal{F}_s = \langle \psi^{-} |\rho|\psi^{-} \rangle = 0.941 \pm 0.005$, where $\rho$ is the measured density matrix and $| \psi^{-} \rangle = \left (|eg\rangle - |ge\rangle\right )/\sqrt{2}$ is the reference Bell singlet state.

The adiabatic method uses the variable coupling of each qubit to the transmission line. When qubits $Q_1$ and $Q_2$ are set to the same frequency and couple to the same resonant mode in the channel with strengths $g_1(t)$ and $g_2(t)$, the single-excitation Hamiltonian for the system can be written in the rotating frame as
\begin{linenomath}
	\begin{align}\label{cpl}
		H/\hbar = g_{1}(t) \left (|e0g \rangle \langle g1g| + |g1g \rangle \langle e0g| \right ) + g_{2}(t) \left (|g0e \rangle \langle g1g| + |g1g \rangle \langle g0e| \right ),
	\end{align}
\end{linenomath}
where $|aNb\rangle$ corresponds to $Q_1$ ($Q_2$) in $|a\rangle$ ($|b\rangle$) with $N$ photons in the resonant transmission line mode. This Hamiltonian supports a ``dark'' eigenstate $|D\rangle$ that has no occupancy in the resonant mode,
\begin{linenomath}
	\begin{align}
		|D(t)\rangle = \frac{1}{\sqrt{2}} \left ( \cos\theta(t) |e0g \rangle - \sin \theta(t) |g0e\rangle \right ),
	\end{align}
\end{linenomath}
where the mixing angle $\theta$ is given by $\tan \theta(t) = g_1(t)/g_2(t)$. With $g_1$ set to zero and $g_2$ to its maximum, the dark state is $|D\rangle = |e0g\rangle$, while exchanging the coupling values $g_1 \leftrightarrow g_2$ yields the dark state $|g0e\rangle$. By adiabatically varying the ratio $g_1(t)/g_2(t)$ in time from zero to its maximum, the system will swap the excitation from $Q_1$ to $Q_2$, without populating the lossy intermediate channel \cite{Wang2012, Vitanov2017}.

Here, we implement a simple adiabatic scheme \cite{Wang2012, Wang2017}, where we vary the couplings in time according to $g_1(t) = \bar{g}\sin \left (\pi t/2t_f\right )$ and $g_2(t) = \bar{g}\cos\left (\pi t/2t_f \right )$. We choose the parameters $\bar{g}/2\pi=15$ MHz and $t_f=132$ ns, minimizing the impact of finite qubit coherence while maintaining sufficient adiabaticity (see \cite{SupplementaryMaterial}). We note that the adiabatic protocol supports better than $90\%$ transfer efficiency even when $\bar{g} = 0.4~\omega_{\rm FSR}$; see \cite{SupplementaryMaterial}.

In Fig.~\ref{fig:adb}a, we demonstrate deterministic adiabatic state transfer from $Q_1$ to $Q_2$. With $Q_1$ in $|e\rangle$ and $Q_1$ and $Q_2$ set on-resonance with a single mode in the channel, we adjust the couplers $G_1$ and $G_2$ adiabatically to complete the state transfer. We show the excited state population of each qubit as a function of time $t$, measured with the resonant mode loss at its intrinsic minimum. We observe the expected gradual population transfer from $Q_1$ to $Q_2$, with $Q_2$'s population reaching its maximum at $t=t_f$, with a transfer efficiency $\eta = P_{e,Q_2}(t=t_f)/P_{e,Q_1}(t=0) = 0.99 \pm 0.01$. We further characterize the state transfer by carrying out quantum process tomography \cite{Neeley2008}, yielding the process matrix $\chi$ shown inset in Fig.~\ref{fig:adb}a, with a process fidelity $\mathcal{F}_{p} = 0.96 \pm 0.01$, limited by qubit decoherence. The process matrix calculated from a master equation simulation displays a small trace distance to the measured $\chi$ matrix of ${\mathcal D} = \sqrt{\mathrm{Tr}\left ([\chi -\chi_{\mathrm{sim}}]^2 \right )} = 0.02 \pm 0.01$, indicating excellent agreement with experiment.

The adiabatic protocol can also be used to generate remote entanglement between $Q_1$ and $Q_2$. With $Q_1$ prepared in $|e\rangle$, we share half its excitation with $Q_2$ using the adiabatic protocol, by stopping the transfer at its midpoint $t=t_f/2$. This generates a Bell singlet state $| \psi^{-} \rangle = \left (|eg\rangle - |ge\rangle\right )/\sqrt{2}$. The qubit excited state population is shown as function of time $t$ in Fig.~\ref{fig:adb}b. We further characterize the Bell state by quantum state tomography \cite{Steffen2006,Neeley2010}, and the reconstructed density matrix $\rho$ is shown inset in Fig.~\ref{fig:adb}b. We find a Bell state fidelity $\mathcal{F}_s = \langle \psi^{-} |\rho|\psi^{-} \rangle = 0.964 \pm 0.007$, referenced to the ideal Bell singlet state $\psi^{-}$, and a concurrence $\mathcal{C} = 0.95 \pm 0.01$ (see \cite{SupplementaryMaterial}). The density matrix $\rho_{\mathrm{sim}}$ calculated from a master equation simulation shows a small trace distance to the measured $\rho$, $\sqrt{\text{Tr}(|\rho-\rho_{\mathrm{sim}}|^2)} = 0.01$, indicating excellent agreement with experiment.

\begin{figure}[t]
	\centering
	\includegraphics[width=8.6cm]{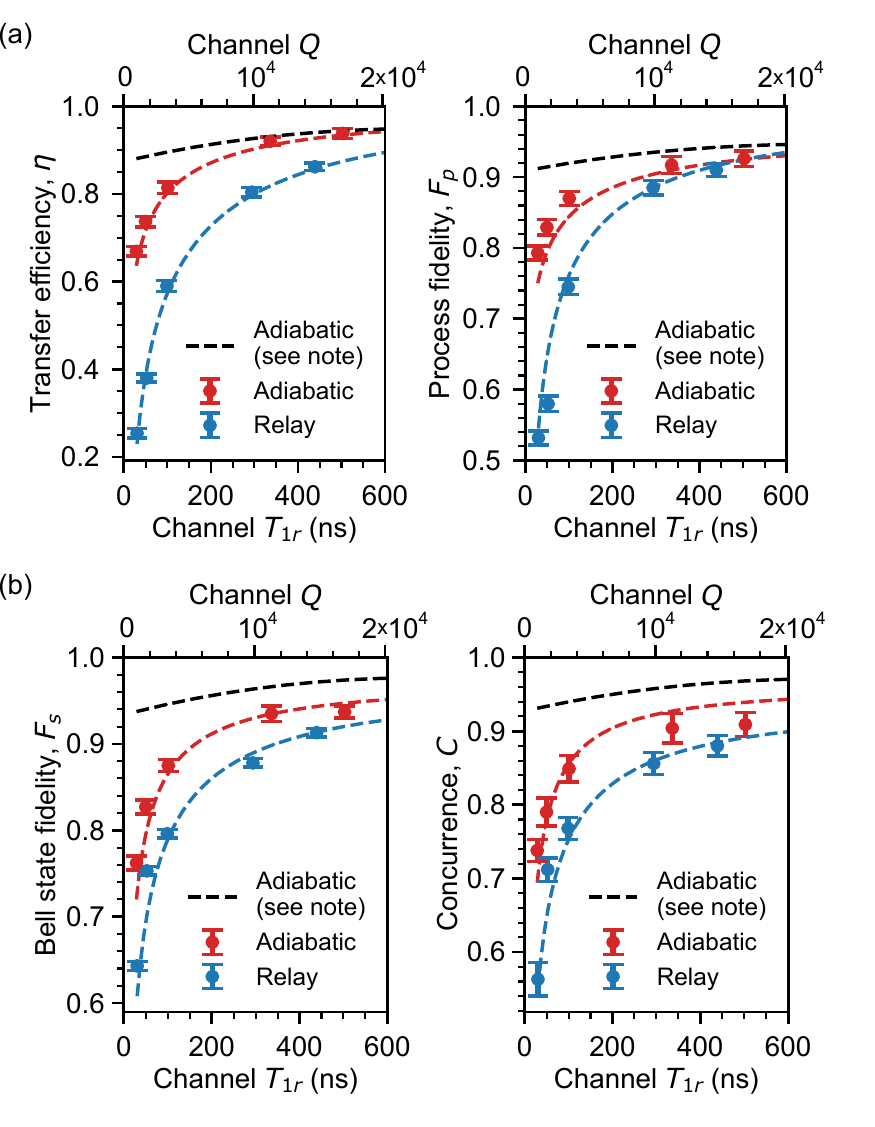}
	\caption{
		\label{fig:dissipation}
		Quantum communication in the presence of channel loss, using both the relay method and adiabatic protocol.
        (a) Measured transfer efficiency $\eta$ (left) and process fidelity $\mathcal{F}_p$ (right) for the adiabatic protocol (red) and the relay method (blue), for different resonant channel mode lifetimes $T_{1r}$, with equivalent quality factors $Q$ shown on top.
        (b) Measured Bell state fidelity $\mathcal{F}_s$ (left) and concurrence $\mathcal{C}$ (right) for adiabatic protocol (red) and relay method (blue). In all panels, error bars are one standard deviation; red and blue dashed lines are from simulations including all sources of loss and black dashed lines are from a master equation simulation for the adiabatic protocol with no $Q_1$ spurious coupling loss (see \cite{SupplementaryMaterial}).
	}
	\centering
\end{figure}

%%%%% FIG 4 PART
We explore the impact of loss on both the relay method and the adiabatic protocol, with results shown as a function of the resonant channel mode energy lifetime $T_{1r}$ in Fig~\ref{fig:dissipation}. For the highest level of dissipation, with $T_{1r} = 28.7$ ns, we measure an adiabatic transfer efficiency $\eta = 0.67 \pm 0.01$, even though the transfer time $t_f$ is four times the resonant mode lifetime. The efficiency is primarily limited by loss in qubit $Q_1$ due to its spurious coupling loss through $D_1$ to the $50~\Omega$ load (see \cite{SupplementaryMaterial}), in good agreement with master equation simulations. Results from a simulation without the spurious coupling are plotted as black dashed lines in Fig~\ref{fig:dissipation}a, limited by a small channel occupation due to the finite adiabaticity of the sequence. We compare these results to the relay method, where we use a weak coupling $|g_{1,2}|/2\pi=5.0$~MHz to ensure the qubits only couple to a single transmission line mode; this results in a total transfer time $2\tau_{\mathrm{swap}} = 100$~ns. We find the adiabatic protocol consistently performs better than the relay method, with a $2.6 \times$ higher transfer efficiency $\eta$ ($2.3 \times$ reduction in transfer loss) and $1.5\times$ higher process fidelity $\mathcal{F}_{p}$ ($2.3 \times$ reduction in process infidelity) compared to the relay method in the most dissipative case; the adiabatic protocol is primarily limited by spurious coupling loss in $Q_1$, while the relay method is limited by loss in the channel (see \cite{SupplementaryMaterial}).

In Fig.~\ref{fig:dissipation}b, we display the entanglement fidelity using the adiabatic protocol with different levels of channel loss, and compare to the relay method. The adiabatic protocol outperforms the relay method in all levels of dissipation. At the highest loss level, where $T_{1r} = 28.7$ ns, the adiabatic protocol achieves $1.2 \times$ higher Bell state fidelity  $\mathcal{F}_s$ ($1.5 \times$ reduction in Bell state infidelity) and $1.3 \times$ higher concurrence $\mathcal{C}$ ($1.7 \times$ reduction in concurrence infidelity) compared to the relay method; the spurious-coupling-free simulation result for the adiabatic protocol is shown by the black dashed lines, limited by a small channel occupation due to the finite adiabaticity of the sequence.

%%%%% CONCLUSION
In conclusion, we describe a unique experimental system in which we can explore the performance of quantum communication protocols in the presence of controllable communication loss. We demonstrate an adiabatic protocol that realizes high-fidelity transfer of quantum states and entangled Bell states, limited mostly by spurious coupling of one qubit to the controlled transmission line loss. The platform we have developed is well-suited to explore the impact of channel loss on other error-protecting quantum communication protocols, such as heralding \cite{Mosley2008,Azuma2015,Kurpiers2019} and entanglement distillation\cite{Kwiat2001,Dong2008,Takahashi2010}. The ability to introduce controlled loss dynamically into the system opens the door to study dissipative dynamics in non-equilibrium systems, enabling approaches such as reservoir engineering \cite{Poyatos1996,Plenio2002}. The adiabatic protocol demonstrated here is applicable to other quantum communication systems, for example phonon-based systems where the communication channel is significantly more lossy \cite{Hermelin2011,McNeil2011,Bienfait2019}. Future demonstrations could employ more advanced adiabatic protocols such as shortcuts to adiabaticity \cite{Baksic2016,Zhou2017} and composite adiabatic passage \cite{Torosov2011,Bruns2018} to further improve fidelity.

\clearpage

\begin{figure}[t]
	\centering
	\includegraphics[width=8.6cm]{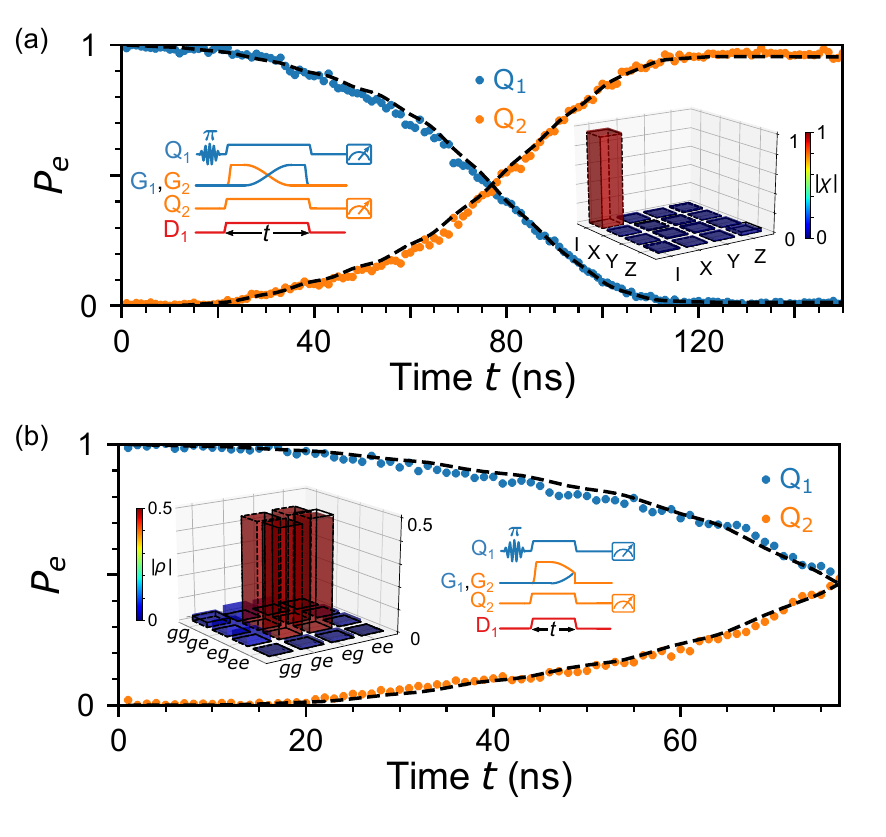}
	\caption{
		\label{fig:adb}
		Quantum state transfer and remote entanglement using the adiabatic protocol.
		(a) Adiabatic state transfer between qubits $Q_1$ and $Q_2$, measured with intrinsic loss in the transmission line. Blue (orange) circles represent excited state populations of $Q_1$ ($Q_2$) measured simultaneously at time $t$.
		Left inset: Control pulse sequence. The couplers are set so that coupling $g_2$ starts at its maximum with $g_1$ set to zero. Dissipation in the resonant channel mode is controlled using $D_1$, here set to zero coupling.
		Right inset: Quantum process tomography, yielding a process fidelity $\mathcal{F}_p=0.96 \pm 0.01$.
		(b) Adiabatic remote entanglement. Right inset shows control pulse sequence: With $Q_1$ initially prepared in $|e\rangle$,  $G_1$ and $G_2$ are controlled using the adiabatic protocol to share half of $Q_1$'s excitation with $Q_2$, resulting in a Bell singlet state $| \psi^{-} \rangle = \left (|eg\rangle - |ge\rangle\right )/\sqrt{2}$. Blue (orange) circles represent excited state populations of $Q_1$ ($Q_2$) measured simultaneously at time $t$. Left inset: Reconstructed density matrix of the final Bell state, yielding a state fidelity $\mathcal{F}_s=0.964 \pm 0.007$ and concurrence $\mathcal{C} = 0.95 \pm 0.01$. In all panels, dashed lines are from master equation simulations accounting for channel dissipation and qubit imperfections (see \cite{SupplementaryMaterial}).}
	\centering
\end{figure}

\clearpage
%%%%% END MATTER
\begin{acknowledgments}
The authors thank A. A. Clerk, P. J. Duda, and B. B. Zhou for helpful discussions. We thank W. D. Oliver and G. Calusine at MIT Lincoln Lab for providing the traveling-wave parametric amplifier (TWPA) used in this work. Devices and experiments were supported by the Air Force Office of Scientific Research and the Army Research Laboratory. K.J.S. was supported by NSF GRFP (NSF DGE-1144085), \'E.D. was supported by LDRD funds from Argonne National Laboratory; A.N.C. was supported in part by the DOE, Office of Basic Energy Sciences. This work was partially supported by the UChicago MRSEC (NSF DMR-1420709) and made use of the Pritzker Nanofabrication Facility, which receives support from SHyNE, a node of the National Science Foundation's National Nanotechnology Coordinated Infrastructure (NSF NNCI-1542205). The authors declare no competing financial interests. Correspondence and requests for materials should be addressed to A. N. Cleland (anc@uchicago.edu).
\end{acknowledgments}

\bibliography{bibliography}

\begin{thebibliography}{37}%
\makeatletter
\providecommand \@ifxundefined [1]{%
 \@ifx{#1\undefined}
}%
\providecommand \@ifnum [1]{%
 \ifnum #1\expandafter \@firstoftwo
 \else \expandafter \@secondoftwo
 \fi
}%
\providecommand \@ifx [1]{%
 \ifx #1\expandafter \@firstoftwo
 \else \expandafter \@secondoftwo
 \fi
}%
\providecommand \natexlab [1]{#1}%
\providecommand \enquote  [1]{``#1''}%
\providecommand \bibnamefont  [1]{#1}%
\providecommand \bibfnamefont [1]{#1}%
\providecommand \citenamefont [1]{#1}%
\providecommand \href@noop [0]{\@secondoftwo}%
\providecommand \href [0]{\begingroup \@sanitize@url \@href}%
\providecommand \@href[1]{\@@startlink{#1}\@@href}%
\providecommand \@@href[1]{\endgroup#1\@@endlink}%
\providecommand \@sanitize@url [0]{\catcode `\\12\catcode `\$12\catcode
  `\&12\catcode `\#12\catcode `\^12\catcode `\_12\catcode `\%12\relax}%
\providecommand \@@startlink[1]{}%
\providecommand \@@endlink[0]{}%
\providecommand \url  [0]{\begingroup\@sanitize@url \@url }%
\providecommand \@url [1]{\endgroup\@href {#1}{\urlprefix }}%
\providecommand \urlprefix  [0]{URL }%
\providecommand \Eprint [0]{\href }%
\providecommand \doibase [0]{http://dx.doi.org/}%
\providecommand \selectlanguage [0]{\@gobble}%
\providecommand \bibinfo  [0]{\@secondoftwo}%
\providecommand \bibfield  [0]{\@secondoftwo}%
\providecommand \translation [1]{[#1]}%
\providecommand \BibitemOpen [0]{}%
\providecommand \bibitemStop [0]{}%
\providecommand \bibitemNoStop [0]{.\EOS\space}%
\providecommand \EOS [0]{\spacefactor3000\relax}%
\providecommand \BibitemShut  [1]{\csname bibitem#1\endcsname}%
\let\auto@bib@innerbib\@empty
%</preamble>
\bibitem [{\citenamefont {Kurpiers}\ \emph {et~al.}(2018)\citenamefont
  {Kurpiers}, \citenamefont {Magnard}, \citenamefont {Walter}, \citenamefont
  {Royer}, \citenamefont {Pechal}, \citenamefont {Heinsoo}, \citenamefont
  {Salathe}, \citenamefont {Akin}, \citenamefont {Storz}, \citenamefont
  {Besse}, \citenamefont {Gasparinetti}, \citenamefont {Blais},\ and\
  \citenamefont {Wallraff}}]{Kurpiers2018}%
  \BibitemOpen
  \bibfield  {author} {\bibinfo {author} {\bibfnamefont {P.}~\bibnamefont
  {Kurpiers}}, \bibinfo {author} {\bibfnamefont {P.}~\bibnamefont {Magnard}},
  \bibinfo {author} {\bibfnamefont {T.}~\bibnamefont {Walter}}, \bibinfo
  {author} {\bibfnamefont {B.}~\bibnamefont {Royer}}, \bibinfo {author}
  {\bibfnamefont {M.}~\bibnamefont {Pechal}}, \bibinfo {author} {\bibfnamefont
  {J.}~\bibnamefont {Heinsoo}}, \bibinfo {author} {\bibfnamefont
  {Y.}~\bibnamefont {Salathe}}, \bibinfo {author} {\bibfnamefont
  {A.}~\bibnamefont {Akin}}, \bibinfo {author} {\bibfnamefont {S.}~\bibnamefont
  {Storz}}, \bibinfo {author} {\bibfnamefont {J.-C.}\ \bibnamefont {Besse}},
  \bibinfo {author} {\bibfnamefont {S.}~\bibnamefont {Gasparinetti}}, \bibinfo
  {author} {\bibfnamefont {A.}~\bibnamefont {Blais}}, \ and\ \bibinfo {author}
  {\bibfnamefont {A.}~\bibnamefont {Wallraff}},\ }\href {\doibase
  10.1038/s41586-018-0195-y} {\bibfield  {journal} {\bibinfo  {journal}
  {Nature}\ }\textbf {\bibinfo {volume} {558}},\ \bibinfo {pages} {264}
  (\bibinfo {year} {2018})}\BibitemShut {NoStop}%
\bibitem [{\citenamefont {Axline}\ \emph {et~al.}(2018)\citenamefont {Axline},
  \citenamefont {Burkhart}, \citenamefont {Pfaff}, \citenamefont {Zhang},
  \citenamefont {Chou}, \citenamefont {Campagne-Ibarcq}, \citenamefont
  {Reinhold}, \citenamefont {Frunzio}, \citenamefont {Girvin}, \citenamefont
  {Jiang}, \citenamefont {Devoret},\ and\ \citenamefont
  {Schoelkopf}}]{Axline2018}%
  \BibitemOpen
  \bibfield  {author} {\bibinfo {author} {\bibfnamefont {C.~J.}\ \bibnamefont
  {Axline}}, \bibinfo {author} {\bibfnamefont {L.~D.}\ \bibnamefont
  {Burkhart}}, \bibinfo {author} {\bibfnamefont {W.}~\bibnamefont {Pfaff}},
  \bibinfo {author} {\bibfnamefont {M.}~\bibnamefont {Zhang}}, \bibinfo
  {author} {\bibfnamefont {K.}~\bibnamefont {Chou}}, \bibinfo {author}
  {\bibfnamefont {P.}~\bibnamefont {Campagne-Ibarcq}}, \bibinfo {author}
  {\bibfnamefont {P.}~\bibnamefont {Reinhold}}, \bibinfo {author}
  {\bibfnamefont {L.}~\bibnamefont {Frunzio}}, \bibinfo {author} {\bibfnamefont
  {S.~M.}\ \bibnamefont {Girvin}}, \bibinfo {author} {\bibfnamefont
  {L.}~\bibnamefont {Jiang}}, \bibinfo {author} {\bibfnamefont {M.~H.}\
  \bibnamefont {Devoret}}, \ and\ \bibinfo {author} {\bibfnamefont {R.~J.}\
  \bibnamefont {Schoelkopf}},\ }\href {\doibase 10.1038/s41567-018-0115-y}
  {\bibfield  {journal} {\bibinfo  {journal} {Nature Physics}\ }\textbf
  {\bibinfo {volume} {14}},\ \bibinfo {pages} {705} (\bibinfo {year}
  {2018})}\BibitemShut {NoStop}%
\bibitem [{\citenamefont {Campagne-Ibarcq}\ \emph {et~al.}(2018)\citenamefont
  {Campagne-Ibarcq}, \citenamefont {Zalys-Geller}, \citenamefont {Narla},
  \citenamefont {Shankar}, \citenamefont {Reinhold}, \citenamefont {Burkhart},
  \citenamefont {Axline}, \citenamefont {Pfaff}, \citenamefont {Frunzio},
  \citenamefont {Schoelkopf},\ and\ \citenamefont {Devoret}}]{Campagne2018}%
  \BibitemOpen
  \bibfield  {author} {\bibinfo {author} {\bibfnamefont {P.}~\bibnamefont
  {Campagne-Ibarcq}}, \bibinfo {author} {\bibfnamefont {E.}~\bibnamefont
  {Zalys-Geller}}, \bibinfo {author} {\bibfnamefont {A.}~\bibnamefont {Narla}},
  \bibinfo {author} {\bibfnamefont {S.}~\bibnamefont {Shankar}}, \bibinfo
  {author} {\bibfnamefont {P.}~\bibnamefont {Reinhold}}, \bibinfo {author}
  {\bibfnamefont {L.}~\bibnamefont {Burkhart}}, \bibinfo {author}
  {\bibfnamefont {C.}~\bibnamefont {Axline}}, \bibinfo {author} {\bibfnamefont
  {W.}~\bibnamefont {Pfaff}}, \bibinfo {author} {\bibfnamefont
  {L.}~\bibnamefont {Frunzio}}, \bibinfo {author} {\bibfnamefont {R.~J.}\
  \bibnamefont {Schoelkopf}}, \ and\ \bibinfo {author} {\bibfnamefont {M.~H.}\
  \bibnamefont {Devoret}},\ }\href {\doibase 10.1103/PhysRevLett.120.200501}
  {\bibfield  {journal} {\bibinfo  {journal} {Physical Review Letters}\
  }\textbf {\bibinfo {volume} {120}},\ \bibinfo {pages} {200501} (\bibinfo
  {year} {2018})}\BibitemShut {NoStop}%
\bibitem [{\citenamefont {Leung}\ \emph {et~al.}(2019)\citenamefont {Leung},
  \citenamefont {Lu}, \citenamefont {Chakram}, \citenamefont {Naik},
  \citenamefont {Earnest}, \citenamefont {Ma}, \citenamefont {Jacobs},
  \citenamefont {Cleland},\ and\ \citenamefont {Schuster}}]{Leung2019}%
  \BibitemOpen
  \bibfield  {author} {\bibinfo {author} {\bibfnamefont {N.}~\bibnamefont
  {Leung}}, \bibinfo {author} {\bibfnamefont {Y.}~\bibnamefont {Lu}}, \bibinfo
  {author} {\bibfnamefont {S.}~\bibnamefont {Chakram}}, \bibinfo {author}
  {\bibfnamefont {R.~K.}\ \bibnamefont {Naik}}, \bibinfo {author}
  {\bibfnamefont {N.}~\bibnamefont {Earnest}}, \bibinfo {author} {\bibfnamefont
  {R.}~\bibnamefont {Ma}}, \bibinfo {author} {\bibfnamefont {K.}~\bibnamefont
  {Jacobs}}, \bibinfo {author} {\bibfnamefont {A.~N.}\ \bibnamefont {Cleland}},
  \ and\ \bibinfo {author} {\bibfnamefont {D.~I.}\ \bibnamefont {Schuster}},\
  }\href {\doibase 10.1038/s41534-019-0128-0} {\bibfield  {journal} {\bibinfo
  {journal} {npj Quantum Information}\ }\textbf {\bibinfo {volume} {5}},\
  \bibinfo {pages} {18} (\bibinfo {year} {2019})}\BibitemShut {NoStop}%
\bibitem [{\citenamefont {Zhong}\ \emph {et~al.}(2019)\citenamefont {Zhong},
  \citenamefont {Chang}, \citenamefont {Satzinger}, \citenamefont {Chou},
  \citenamefont {Bienfait}, \citenamefont {Conner}, \citenamefont {Dumur},
  \citenamefont {Grebel}, \citenamefont {Peairs}, \citenamefont {Povey},
  \citenamefont {Schuster},\ and\ \citenamefont {Cleland}}]{Zhong2019}%
  \BibitemOpen
  \bibfield  {author} {\bibinfo {author} {\bibfnamefont {Y.~P.}\ \bibnamefont
  {Zhong}}, \bibinfo {author} {\bibfnamefont {H.-S.}\ \bibnamefont {Chang}},
  \bibinfo {author} {\bibfnamefont {K.~J.}\ \bibnamefont {Satzinger}}, \bibinfo
  {author} {\bibfnamefont {M.-H.}\ \bibnamefont {Chou}}, \bibinfo {author}
  {\bibfnamefont {A.}~\bibnamefont {Bienfait}}, \bibinfo {author}
  {\bibfnamefont {C.~R.}\ \bibnamefont {Conner}}, \bibinfo {author}
  {\bibfnamefont {{\'E}.}~\bibnamefont {Dumur}}, \bibinfo {author}
  {\bibfnamefont {J.}~\bibnamefont {Grebel}}, \bibinfo {author} {\bibfnamefont
  {G.~A.}\ \bibnamefont {Peairs}}, \bibinfo {author} {\bibfnamefont {R.~G.}\
  \bibnamefont {Povey}}, \bibinfo {author} {\bibfnamefont {D.~I.}\ \bibnamefont
  {Schuster}}, \ and\ \bibinfo {author} {\bibfnamefont {A.~N.}\ \bibnamefont
  {Cleland}},\ }\href {\doibase 10.1038/s41567-019-0507-7} {\bibfield
  {journal} {\bibinfo  {journal} {Nature Physics}\ }\textbf {\bibinfo {volume}
  {15}},\ \bibinfo {pages} {741} (\bibinfo {year} {2019})}\BibitemShut
  {NoStop}%
\bibitem [{\citenamefont {Bienfait}\ \emph {et~al.}(2019)\citenamefont
  {Bienfait}, \citenamefont {Satzinger}, \citenamefont {Zhong}, \citenamefont
  {Chang}, \citenamefont {Chou}, \citenamefont {Conner}, \citenamefont {Dumur},
  \citenamefont {Grebel}, \citenamefont {Peairs}, \citenamefont {Povey},\ and\
  \citenamefont {Cleland}}]{Bienfait2019}%
  \BibitemOpen
  \bibfield  {author} {\bibinfo {author} {\bibfnamefont {A.}~\bibnamefont
  {Bienfait}}, \bibinfo {author} {\bibfnamefont {K.~J.}\ \bibnamefont
  {Satzinger}}, \bibinfo {author} {\bibfnamefont {Y.~P.}\ \bibnamefont
  {Zhong}}, \bibinfo {author} {\bibfnamefont {H.-S.}\ \bibnamefont {Chang}},
  \bibinfo {author} {\bibfnamefont {M.-H.}\ \bibnamefont {Chou}}, \bibinfo
  {author} {\bibfnamefont {C.~R.}\ \bibnamefont {Conner}}, \bibinfo {author}
  {\bibfnamefont {{\'E}.}~\bibnamefont {Dumur}}, \bibinfo {author}
  {\bibfnamefont {J.}~\bibnamefont {Grebel}}, \bibinfo {author} {\bibfnamefont
  {G.~A.}\ \bibnamefont {Peairs}}, \bibinfo {author} {\bibfnamefont {R.~G.}\
  \bibnamefont {Povey}}, \ and\ \bibinfo {author} {\bibfnamefont {A.~N.}\
  \bibnamefont {Cleland}},\ }\href {\doibase 10.1126/science.aaw8415}
  {\bibfield  {journal} {\bibinfo  {journal} {Science}\ }\textbf {\bibinfo
  {volume} {364}},\ \bibinfo {pages} {368} (\bibinfo {year}
  {2019})}\BibitemShut {NoStop}%
\bibitem [{\citenamefont {Vitanov}\ \emph {et~al.}(2017)\citenamefont
  {Vitanov}, \citenamefont {Rangelov}, \citenamefont {Shore},\ and\
  \citenamefont {Bergmann}}]{Vitanov2017}%
  \BibitemOpen
  \bibfield  {author} {\bibinfo {author} {\bibfnamefont {N.~V.}\ \bibnamefont
  {Vitanov}}, \bibinfo {author} {\bibfnamefont {A.~A.}\ \bibnamefont
  {Rangelov}}, \bibinfo {author} {\bibfnamefont {B.~W.}\ \bibnamefont {Shore}},
  \ and\ \bibinfo {author} {\bibfnamefont {K.}~\bibnamefont {Bergmann}},\
  }\href {\doibase 10.1103/RevModPhys.89.015006} {\bibfield  {journal}
  {\bibinfo  {journal} {Reviews of Modern Physics}\ }\textbf {\bibinfo {volume}
  {89}},\ \bibinfo {pages} {015006} (\bibinfo {year} {2017})}\BibitemShut
  {NoStop}%
\bibitem [{\citenamefont {Bergmann}\ \emph {et~al.}(2019)\citenamefont
  {Bergmann}, \citenamefont {Nagerl}, \citenamefont {Panda}, \citenamefont
  {Gabrielse}, \citenamefont {Miloglyadov}, \citenamefont {Quack},
  \citenamefont {Seyfang}, \citenamefont {Wichmann}, \citenamefont {Ospelkaus},
  \citenamefont {Kuhn}, \citenamefont {Longhi}, \citenamefont {Szameit},
  \citenamefont {Pirro}, \citenamefont {Hillebrands}, \citenamefont {Zhu},
  \citenamefont {Zhu}, \citenamefont {Drewsen}, \citenamefont {Hensinger},
  \citenamefont {Weidt}, \citenamefont {Halfmann}, \citenamefont {Wang},
  \citenamefont {Paraoanu}, \citenamefont {Vitanov}, \citenamefont {Mompart},
  \citenamefont {Busch}, \citenamefont {Barnum}, \citenamefont {Grimes},
  \citenamefont {Field}, \citenamefont {Raizen}, \citenamefont {Narevicius},
  \citenamefont {Auzinsh}, \citenamefont {Budker}, \citenamefont {Pálffy},\
  and\ \citenamefont {Keitel}}]{Bergmann2019}%
  \BibitemOpen
  \bibfield  {author} {\bibinfo {author} {\bibfnamefont {K.}~\bibnamefont
  {Bergmann}}, \bibinfo {author} {\bibfnamefont {H.-C.}\ \bibnamefont
  {Nagerl}}, \bibinfo {author} {\bibfnamefont {C.}~\bibnamefont {Panda}},
  \bibinfo {author} {\bibfnamefont {G.}~\bibnamefont {Gabrielse}}, \bibinfo
  {author} {\bibfnamefont {E.}~\bibnamefont {Miloglyadov}}, \bibinfo {author}
  {\bibfnamefont {M.}~\bibnamefont {Quack}}, \bibinfo {author} {\bibfnamefont
  {G.}~\bibnamefont {Seyfang}}, \bibinfo {author} {\bibfnamefont
  {G.}~\bibnamefont {Wichmann}}, \bibinfo {author} {\bibfnamefont
  {S.}~\bibnamefont {Ospelkaus}}, \bibinfo {author} {\bibfnamefont
  {A.}~\bibnamefont {Kuhn}}, \bibinfo {author} {\bibfnamefont {S.}~\bibnamefont
  {Longhi}}, \bibinfo {author} {\bibfnamefont {A.}~\bibnamefont {Szameit}},
  \bibinfo {author} {\bibfnamefont {P.}~\bibnamefont {Pirro}}, \bibinfo
  {author} {\bibfnamefont {B.}~\bibnamefont {Hillebrands}}, \bibinfo {author}
  {\bibfnamefont {X.-F.}\ \bibnamefont {Zhu}}, \bibinfo {author} {\bibfnamefont
  {J.}~\bibnamefont {Zhu}}, \bibinfo {author} {\bibfnamefont {M.}~\bibnamefont
  {Drewsen}}, \bibinfo {author} {\bibfnamefont {W.~K.}\ \bibnamefont
  {Hensinger}}, \bibinfo {author} {\bibfnamefont {S.}~\bibnamefont {Weidt}},
  \bibinfo {author} {\bibfnamefont {T.}~\bibnamefont {Halfmann}}, \bibinfo
  {author} {\bibfnamefont {H.-L.}\ \bibnamefont {Wang}}, \bibinfo {author}
  {\bibfnamefont {G.~S.}\ \bibnamefont {Paraoanu}}, \bibinfo {author}
  {\bibfnamefont {N.~V.}\ \bibnamefont {Vitanov}}, \bibinfo {author}
  {\bibfnamefont {J.}~\bibnamefont {Mompart}}, \bibinfo {author} {\bibfnamefont
  {T.}~\bibnamefont {Busch}}, \bibinfo {author} {\bibfnamefont {T.~J.}\
  \bibnamefont {Barnum}}, \bibinfo {author} {\bibfnamefont {D.~D.}\
  \bibnamefont {Grimes}}, \bibinfo {author} {\bibfnamefont {R.~W.}\
  \bibnamefont {Field}}, \bibinfo {author} {\bibfnamefont {M.~G.}\ \bibnamefont
  {Raizen}}, \bibinfo {author} {\bibfnamefont {E.}~\bibnamefont {Narevicius}},
  \bibinfo {author} {\bibfnamefont {M.}~\bibnamefont {Auzinsh}}, \bibinfo
  {author} {\bibfnamefont {D.}~\bibnamefont {Budker}}, \bibinfo {author}
  {\bibfnamefont {A.}~\bibnamefont {Pálffy}}, \ and\ \bibinfo {author}
  {\bibfnamefont {C.~H.}\ \bibnamefont {Keitel}},\ }\href {\doibase
  10.1088/1361-6455/ab3995} {\bibfield  {journal} {\bibinfo  {journal} {Journal
  of Physics B: Atomic, Molecular and Optical Physics}\ }\textbf {\bibinfo
  {volume} {52}},\ \bibinfo {pages} {202001} (\bibinfo {year}
  {2019})}\BibitemShut {NoStop}%
\bibitem [{\citenamefont {Sillanpaa}\ \emph {et~al.}(2007)\citenamefont
  {Sillanpaa}, \citenamefont {Park},\ and\ \citenamefont
  {Simmonds}}]{Sillanpaa2007}%
  \BibitemOpen
  \bibfield  {author} {\bibinfo {author} {\bibfnamefont {M.~A.}\ \bibnamefont
  {Sillanpaa}}, \bibinfo {author} {\bibfnamefont {J.~I.}\ \bibnamefont {Park}},
  \ and\ \bibinfo {author} {\bibfnamefont {R.~W.}\ \bibnamefont {Simmonds}},\
  }\href {\doibase 10.1038/nature06124} {\bibfield  {journal} {\bibinfo
  {journal} {Nature}\ }\textbf {\bibinfo {volume} {449}},\ \bibinfo {pages}
  {438} (\bibinfo {year} {2007})}\BibitemShut {NoStop}%
\bibitem [{\citenamefont {Ansmann}\ \emph {et~al.}(2009)\citenamefont
  {Ansmann}, \citenamefont {Wang}, \citenamefont {Bialczak}, \citenamefont
  {Hofheinz}, \citenamefont {Lucero}, \citenamefont {Neeley}, \citenamefont
  {O'Connell}, \citenamefont {Sank}, \citenamefont {Weides}, \citenamefont
  {Wenner}, \citenamefont {Cleland},\ and\ \citenamefont
  {Martinis}}]{Ansmann2009}%
  \BibitemOpen
  \bibfield  {author} {\bibinfo {author} {\bibfnamefont {M.}~\bibnamefont
  {Ansmann}}, \bibinfo {author} {\bibfnamefont {H.}~\bibnamefont {Wang}},
  \bibinfo {author} {\bibfnamefont {R.~C.}\ \bibnamefont {Bialczak}}, \bibinfo
  {author} {\bibfnamefont {M.}~\bibnamefont {Hofheinz}}, \bibinfo {author}
  {\bibfnamefont {E.}~\bibnamefont {Lucero}}, \bibinfo {author} {\bibfnamefont
  {M.}~\bibnamefont {Neeley}}, \bibinfo {author} {\bibfnamefont {A.~D.}\
  \bibnamefont {O'Connell}}, \bibinfo {author} {\bibfnamefont {D.}~\bibnamefont
  {Sank}}, \bibinfo {author} {\bibfnamefont {M.}~\bibnamefont {Weides}},
  \bibinfo {author} {\bibfnamefont {J.}~\bibnamefont {Wenner}}, \bibinfo
  {author} {\bibfnamefont {A.~N.}\ \bibnamefont {Cleland}}, \ and\ \bibinfo
  {author} {\bibfnamefont {J.~M.}\ \bibnamefont {Martinis}},\ }\href {\doibase
  10.1038/nature08363} {\bibfield  {journal} {\bibinfo  {journal} {Nature}\
  }\textbf {\bibinfo {volume} {461}},\ \bibinfo {pages} {504} (\bibinfo {year}
  {2009})}\BibitemShut {NoStop}%
\bibitem [{\citenamefont {Koch}\ \emph {et~al.}(2007)\citenamefont {Koch},
  \citenamefont {Yu}, \citenamefont {Gambetta}, \citenamefont {Houck},
  \citenamefont {Schuster}, \citenamefont {Majer}, \citenamefont {Blais},
  \citenamefont {Devoret}, \citenamefont {Girvin},\ and\ \citenamefont
  {Schoelkopf}}]{Koch2007}%
  \BibitemOpen
  \bibfield  {author} {\bibinfo {author} {\bibfnamefont {J.}~\bibnamefont
  {Koch}}, \bibinfo {author} {\bibfnamefont {T.~M.}\ \bibnamefont {Yu}},
  \bibinfo {author} {\bibfnamefont {J.}~\bibnamefont {Gambetta}}, \bibinfo
  {author} {\bibfnamefont {A.~A.}\ \bibnamefont {Houck}}, \bibinfo {author}
  {\bibfnamefont {D.~I.}\ \bibnamefont {Schuster}}, \bibinfo {author}
  {\bibfnamefont {J.}~\bibnamefont {Majer}}, \bibinfo {author} {\bibfnamefont
  {A.}~\bibnamefont {Blais}}, \bibinfo {author} {\bibfnamefont {M.~H.}\
  \bibnamefont {Devoret}}, \bibinfo {author} {\bibfnamefont {S.~M.}\
  \bibnamefont {Girvin}}, \ and\ \bibinfo {author} {\bibfnamefont {R.~J.}\
  \bibnamefont {Schoelkopf}},\ }\href {\doibase 10.1103/PhysRevA.76.042319}
  {\bibfield  {journal} {\bibinfo  {journal} {Physical Review A}\ }\textbf
  {\bibinfo {volume} {76}},\ \bibinfo {pages} {042319} (\bibinfo {year}
  {2007})}\BibitemShut {NoStop}%
\bibitem [{\citenamefont {Barends}\ \emph {et~al.}(2013)\citenamefont
  {Barends}, \citenamefont {Kelly}, \citenamefont {Megrant}, \citenamefont
  {Sank}, \citenamefont {Jeffrey}, \citenamefont {Chen}, \citenamefont {Yin},
  \citenamefont {Chiaro}, \citenamefont {Mutus}, \citenamefont {Neill},
  \citenamefont {O'Malley}, \citenamefont {Roushan}, \citenamefont {Wenner},
  \citenamefont {White}, \citenamefont {Cleland},\ and\ \citenamefont
  {Martinis}}]{Barends2013}%
  \BibitemOpen
  \bibfield  {author} {\bibinfo {author} {\bibfnamefont {R.}~\bibnamefont
  {Barends}}, \bibinfo {author} {\bibfnamefont {J.}~\bibnamefont {Kelly}},
  \bibinfo {author} {\bibfnamefont {A.}~\bibnamefont {Megrant}}, \bibinfo
  {author} {\bibfnamefont {D.}~\bibnamefont {Sank}}, \bibinfo {author}
  {\bibfnamefont {E.}~\bibnamefont {Jeffrey}}, \bibinfo {author} {\bibfnamefont
  {Y.}~\bibnamefont {Chen}}, \bibinfo {author} {\bibfnamefont {Y.}~\bibnamefont
  {Yin}}, \bibinfo {author} {\bibfnamefont {B.}~\bibnamefont {Chiaro}},
  \bibinfo {author} {\bibfnamefont {J.}~\bibnamefont {Mutus}}, \bibinfo
  {author} {\bibfnamefont {C.}~\bibnamefont {Neill}}, \bibinfo {author}
  {\bibfnamefont {P.}~\bibnamefont {O'Malley}}, \bibinfo {author}
  {\bibfnamefont {P.}~\bibnamefont {Roushan}}, \bibinfo {author} {\bibfnamefont
  {J.}~\bibnamefont {Wenner}}, \bibinfo {author} {\bibfnamefont {T.~C.}\
  \bibnamefont {White}}, \bibinfo {author} {\bibfnamefont {A.~N.}\ \bibnamefont
  {Cleland}}, \ and\ \bibinfo {author} {\bibfnamefont {J.~M.}\ \bibnamefont
  {Martinis}},\ }\href {\doibase 10.1103/PhysRevLett.111.080502} {\bibfield
  {journal} {\bibinfo  {journal} {Physical Review Letters}\ }\textbf {\bibinfo
  {volume} {111}},\ \bibinfo {pages} {080502} (\bibinfo {year}
  {2013})}\BibitemShut {NoStop}%
\bibitem [{\citenamefont {Chen}\ \emph {et~al.}(2014)\citenamefont {Chen},
  \citenamefont {Neill}, \citenamefont {Roushan}, \citenamefont {Leung},
  \citenamefont {Fang}, \citenamefont {Barends}, \citenamefont {Kelly},
  \citenamefont {Campbell}, \citenamefont {Chen}, \citenamefont {Chiaro},
  \citenamefont {Dunsworth}, \citenamefont {Jeffrey}, \citenamefont {Megrant},
  \citenamefont {Mutus}, \citenamefont {O'Malley}, \citenamefont {Quintana},
  \citenamefont {Sank}, \citenamefont {Vainsencher}, \citenamefont {Wenner},
  \citenamefont {White}, \citenamefont {Geller}, \citenamefont {Cleland},\ and\
  \citenamefont {Martinis}}]{Chen2014}%
  \BibitemOpen
  \bibfield  {author} {\bibinfo {author} {\bibfnamefont {Y.}~\bibnamefont
  {Chen}}, \bibinfo {author} {\bibfnamefont {C.}~\bibnamefont {Neill}},
  \bibinfo {author} {\bibfnamefont {P.}~\bibnamefont {Roushan}}, \bibinfo
  {author} {\bibfnamefont {N.}~\bibnamefont {Leung}}, \bibinfo {author}
  {\bibfnamefont {M.}~\bibnamefont {Fang}}, \bibinfo {author} {\bibfnamefont
  {R.}~\bibnamefont {Barends}}, \bibinfo {author} {\bibfnamefont
  {J.}~\bibnamefont {Kelly}}, \bibinfo {author} {\bibfnamefont
  {B.}~\bibnamefont {Campbell}}, \bibinfo {author} {\bibfnamefont
  {Z.}~\bibnamefont {Chen}}, \bibinfo {author} {\bibfnamefont {B.}~\bibnamefont
  {Chiaro}}, \bibinfo {author} {\bibfnamefont {A.}~\bibnamefont {Dunsworth}},
  \bibinfo {author} {\bibfnamefont {E.}~\bibnamefont {Jeffrey}}, \bibinfo
  {author} {\bibfnamefont {A.}~\bibnamefont {Megrant}}, \bibinfo {author}
  {\bibfnamefont {J.~Y.}\ \bibnamefont {Mutus}}, \bibinfo {author}
  {\bibfnamefont {P.~J.~J.}\ \bibnamefont {O'Malley}}, \bibinfo {author}
  {\bibfnamefont {C.~M.}\ \bibnamefont {Quintana}}, \bibinfo {author}
  {\bibfnamefont {D.}~\bibnamefont {Sank}}, \bibinfo {author} {\bibfnamefont
  {A.}~\bibnamefont {Vainsencher}}, \bibinfo {author} {\bibfnamefont
  {J.}~\bibnamefont {Wenner}}, \bibinfo {author} {\bibfnamefont {T.~C.}\
  \bibnamefont {White}}, \bibinfo {author} {\bibfnamefont {M.~R.}\ \bibnamefont
  {Geller}}, \bibinfo {author} {\bibfnamefont {A.~N.}\ \bibnamefont {Cleland}},
  \ and\ \bibinfo {author} {\bibfnamefont {J.~M.}\ \bibnamefont {Martinis}},\
  }\href {\doibase 10.1103/PhysRevLett.113.220502} {\bibfield  {journal}
  {\bibinfo  {journal} {Physical Review Letters}\ }\textbf {\bibinfo {volume}
  {113}},\ \bibinfo {pages} {220502} (\bibinfo {year} {2014})}\BibitemShut
  {NoStop}%
\bibitem [{\citenamefont {Schuster}\ \emph {et~al.}(2005)\citenamefont
  {Schuster}, \citenamefont {Wallraff}, \citenamefont {Blais}, \citenamefont
  {Frunzio}, \citenamefont {Huang}, \citenamefont {Majer}, \citenamefont
  {Girvin},\ and\ \citenamefont {Schoelkopf}}]{Schuster2005}%
  \BibitemOpen
  \bibfield  {author} {\bibinfo {author} {\bibfnamefont {D.~I.}\ \bibnamefont
  {Schuster}}, \bibinfo {author} {\bibfnamefont {A.}~\bibnamefont {Wallraff}},
  \bibinfo {author} {\bibfnamefont {A.}~\bibnamefont {Blais}}, \bibinfo
  {author} {\bibfnamefont {L.}~\bibnamefont {Frunzio}}, \bibinfo {author}
  {\bibfnamefont {R.-S.}\ \bibnamefont {Huang}}, \bibinfo {author}
  {\bibfnamefont {J.}~\bibnamefont {Majer}}, \bibinfo {author} {\bibfnamefont
  {S.~M.}\ \bibnamefont {Girvin}}, \ and\ \bibinfo {author} {\bibfnamefont
  {R.~J.}\ \bibnamefont {Schoelkopf}},\ }\href {\doibase
  10.1103/PhysRevLett.94.123602} {\bibfield  {journal} {\bibinfo  {journal}
  {Physical Review Letters}\ }\textbf {\bibinfo {volume} {94}},\ \bibinfo
  {pages} {123602} (\bibinfo {year} {2005})}\BibitemShut {NoStop}%
\bibitem [{\citenamefont {Wallraff}\ \emph {et~al.}(2005)\citenamefont
  {Wallraff}, \citenamefont {Schuster}, \citenamefont {Blais}, \citenamefont
  {Frunzio}, \citenamefont {Majer}, \citenamefont {Devoret}, \citenamefont
  {Girvin},\ and\ \citenamefont {Schoelkopf}}]{Wallraff2005}%
  \BibitemOpen
  \bibfield  {author} {\bibinfo {author} {\bibfnamefont {A.}~\bibnamefont
  {Wallraff}}, \bibinfo {author} {\bibfnamefont {D.~I.}\ \bibnamefont
  {Schuster}}, \bibinfo {author} {\bibfnamefont {A.}~\bibnamefont {Blais}},
  \bibinfo {author} {\bibfnamefont {L.}~\bibnamefont {Frunzio}}, \bibinfo
  {author} {\bibfnamefont {J.}~\bibnamefont {Majer}}, \bibinfo {author}
  {\bibfnamefont {M.~H.}\ \bibnamefont {Devoret}}, \bibinfo {author}
  {\bibfnamefont {S.~M.}\ \bibnamefont {Girvin}}, \ and\ \bibinfo {author}
  {\bibfnamefont {R.~J.}\ \bibnamefont {Schoelkopf}},\ }\href {\doibase
  10.1103/PhysRevLett.95.060501} {\bibfield  {journal} {\bibinfo  {journal}
  {Physical Review Letters}\ }\textbf {\bibinfo {volume} {95}},\ \bibinfo
  {pages} {060501} (\bibinfo {year} {2005})}\BibitemShut {NoStop}%
\bibitem [{\citenamefont {Blais}\ \emph {et~al.}(2004)\citenamefont {Blais},
  \citenamefont {Huang}, \citenamefont {Wallraff}, \citenamefont {Girvin},\
  and\ \citenamefont {Schoelkopf}}]{Blais2004}%
  \BibitemOpen
  \bibfield  {author} {\bibinfo {author} {\bibfnamefont {A.}~\bibnamefont
  {Blais}}, \bibinfo {author} {\bibfnamefont {R.-S.}\ \bibnamefont {Huang}},
  \bibinfo {author} {\bibfnamefont {A.}~\bibnamefont {Wallraff}}, \bibinfo
  {author} {\bibfnamefont {S.~M.}\ \bibnamefont {Girvin}}, \ and\ \bibinfo
  {author} {\bibfnamefont {R.~J.}\ \bibnamefont {Schoelkopf}},\ }\href
  {\doibase 10.1103/PhysRevA.69.062320} {\bibfield  {journal} {\bibinfo
  {journal} {Physical Review A}\ }\textbf {\bibinfo {volume} {69}},\ \bibinfo
  {pages} {062320} (\bibinfo {year} {2004})}\BibitemShut {NoStop}%
\bibitem [{\citenamefont {Chang}\ \emph {et~al.}(2020)\citenamefont {Chang},
  \citenamefont {Zhong}, \citenamefont {Satzinger}, \citenamefont {Chou},
  \citenamefont {Bienfait}, \citenamefont {Conner}, \citenamefont {Dumur},
  \citenamefont {Grebel}, \citenamefont {Peairs}, \citenamefont {Povey},\ and\
  \citenamefont {Cleland}}]{Chang2020}%
  \BibitemOpen
  \bibfield  {author} {\bibinfo {author} {\bibfnamefont {H.-S.}\ \bibnamefont
  {Chang}}, \bibinfo {author} {\bibfnamefont {Y.~P.}\ \bibnamefont {Zhong}},
  \bibinfo {author} {\bibfnamefont {K.~J.}\ \bibnamefont {Satzinger}}, \bibinfo
  {author} {\bibfnamefont {M.-H.}\ \bibnamefont {Chou}}, \bibinfo {author}
  {\bibfnamefont {A.}~\bibnamefont {Bienfait}}, \bibinfo {author}
  {\bibfnamefont {C.~R.}\ \bibnamefont {Conner}}, \bibinfo {author}
  {\bibfnamefont {{\'E}.}~\bibnamefont {Dumur}}, \bibinfo {author}
  {\bibfnamefont {J.}~\bibnamefont {Grebel}}, \bibinfo {author} {\bibfnamefont
  {G.~A.}\ \bibnamefont {Peairs}}, \bibinfo {author} {\bibfnamefont {R.~G.}\
  \bibnamefont {Povey}}, \ and\ \bibinfo {author} {\bibfnamefont {A.~N.}\
  \bibnamefont {Cleland}},\ }\href@noop {} {\bibfield  {journal} {\bibinfo
  {journal} {In preparation}\ } (\bibinfo {year} {2020})}\BibitemShut {NoStop}%
\bibitem [{\citenamefont {Material}()}]{SupplementaryMaterial}%
  \BibitemOpen
  \bibfield  {author} {\bibinfo {author} {\bibfnamefont {S.}~\bibnamefont
  {Material}},\ }\href@noop {} {\ }\BibitemShut {NoStop}%
\bibitem [{\citenamefont {Wang}\ and\ \citenamefont {Clerk}(2012)}]{Wang2012}%
  \BibitemOpen
  \bibfield  {author} {\bibinfo {author} {\bibfnamefont {Y.-D.}\ \bibnamefont
  {Wang}}\ and\ \bibinfo {author} {\bibfnamefont {A.~A.}\ \bibnamefont
  {Clerk}},\ }\href {\doibase 10.1088/1367-2630/14/10/105010} {\bibfield
  {journal} {\bibinfo  {journal} {New Journal of Physics}\ }\textbf {\bibinfo
  {volume} {14}},\ \bibinfo {pages} {105010} (\bibinfo {year}
  {2012})}\BibitemShut {NoStop}%
\bibitem [{\citenamefont {Wang}\ \emph {et~al.}(2017)\citenamefont {Wang},
  \citenamefont {Zhang}, \citenamefont {Yan},\ and\ \citenamefont
  {Chesi}}]{Wang2017}%
  \BibitemOpen
  \bibfield  {author} {\bibinfo {author} {\bibfnamefont {Y.-D.}\ \bibnamefont
  {Wang}}, \bibinfo {author} {\bibfnamefont {R.}~\bibnamefont {Zhang}},
  \bibinfo {author} {\bibfnamefont {X.-B.}\ \bibnamefont {Yan}}, \ and\
  \bibinfo {author} {\bibfnamefont {S.}~\bibnamefont {Chesi}},\ }\href
  {\doibase 10.1088/1367-2630/aa7f5d} {\bibfield  {journal} {\bibinfo
  {journal} {New Journal of Physics}\ }\textbf {\bibinfo {volume} {19}},\
  \bibinfo {pages} {093016} (\bibinfo {year} {2017})}\BibitemShut {NoStop}%
\bibitem [{\citenamefont {Neeley}\ \emph {et~al.}(2008)\citenamefont {Neeley},
  \citenamefont {Ansmann}, \citenamefont {Bialczak}, \citenamefont {Hofheinz},
  \citenamefont {Katz}, \citenamefont {Lucero}, \citenamefont {O'Connell},
  \citenamefont {Wang}, \citenamefont {Cleland},\ and\ \citenamefont
  {Martinis}}]{Neeley2008}%
  \BibitemOpen
  \bibfield  {author} {\bibinfo {author} {\bibfnamefont {M.}~\bibnamefont
  {Neeley}}, \bibinfo {author} {\bibfnamefont {M.}~\bibnamefont {Ansmann}},
  \bibinfo {author} {\bibfnamefont {R.~C.}\ \bibnamefont {Bialczak}}, \bibinfo
  {author} {\bibfnamefont {M.}~\bibnamefont {Hofheinz}}, \bibinfo {author}
  {\bibfnamefont {N.}~\bibnamefont {Katz}}, \bibinfo {author} {\bibfnamefont
  {E.}~\bibnamefont {Lucero}}, \bibinfo {author} {\bibfnamefont
  {A.}~\bibnamefont {O'Connell}}, \bibinfo {author} {\bibfnamefont
  {H.}~\bibnamefont {Wang}}, \bibinfo {author} {\bibfnamefont {A.~N.}\
  \bibnamefont {Cleland}}, \ and\ \bibinfo {author} {\bibfnamefont {J.~M.}\
  \bibnamefont {Martinis}},\ }\href {\doibase 10.1038/nphys972} {\bibfield
  {journal} {\bibinfo  {journal} {Nature Physics}\ }\textbf {\bibinfo {volume}
  {4}},\ \bibinfo {pages} {523} (\bibinfo {year} {2008})}\BibitemShut {NoStop}%
\bibitem [{\citenamefont {Steffen}\ \emph {et~al.}(2006)\citenamefont
  {Steffen}, \citenamefont {Ansmann}, \citenamefont {Bialczak}, \citenamefont
  {Katz}, \citenamefont {Lucero}, \citenamefont {McDermott}, \citenamefont
  {Neeley}, \citenamefont {Weig}, \citenamefont {Cleland},\ and\ \citenamefont
  {Martinis}}]{Steffen2006}%
  \BibitemOpen
  \bibfield  {author} {\bibinfo {author} {\bibfnamefont {M.}~\bibnamefont
  {Steffen}}, \bibinfo {author} {\bibfnamefont {M.}~\bibnamefont {Ansmann}},
  \bibinfo {author} {\bibfnamefont {R.~C.}\ \bibnamefont {Bialczak}}, \bibinfo
  {author} {\bibfnamefont {N.}~\bibnamefont {Katz}}, \bibinfo {author}
  {\bibfnamefont {E.}~\bibnamefont {Lucero}}, \bibinfo {author} {\bibfnamefont
  {R.}~\bibnamefont {McDermott}}, \bibinfo {author} {\bibfnamefont
  {M.}~\bibnamefont {Neeley}}, \bibinfo {author} {\bibfnamefont {E.~M.}\
  \bibnamefont {Weig}}, \bibinfo {author} {\bibfnamefont {A.~N.}\ \bibnamefont
  {Cleland}}, \ and\ \bibinfo {author} {\bibfnamefont {J.~M.}\ \bibnamefont
  {Martinis}},\ }\href {\doibase 10.1126/science.1130886} {\bibfield  {journal}
  {\bibinfo  {journal} {Science}\ }\textbf {\bibinfo {volume} {313}},\ \bibinfo
  {pages} {1423} (\bibinfo {year} {2006})}\BibitemShut {NoStop}%
\bibitem [{\citenamefont {Neeley}\ \emph {et~al.}(2010)\citenamefont {Neeley},
  \citenamefont {Bialczak}, \citenamefont {Lenander}, \citenamefont {Lucero},
  \citenamefont {Mariantoni}, \citenamefont {O'Connell}, \citenamefont {Sank},
  \citenamefont {Wang}, \citenamefont {Weides}, \citenamefont {Wenner},
  \citenamefont {Yin}, \citenamefont {Yamamoto}, \citenamefont {Cleland},\ and\
  \citenamefont {Martinis}}]{Neeley2010}%
  \BibitemOpen
  \bibfield  {author} {\bibinfo {author} {\bibfnamefont {M.}~\bibnamefont
  {Neeley}}, \bibinfo {author} {\bibfnamefont {R.~C.}\ \bibnamefont
  {Bialczak}}, \bibinfo {author} {\bibfnamefont {M.}~\bibnamefont {Lenander}},
  \bibinfo {author} {\bibfnamefont {E.}~\bibnamefont {Lucero}}, \bibinfo
  {author} {\bibfnamefont {M.}~\bibnamefont {Mariantoni}}, \bibinfo {author}
  {\bibfnamefont {A.~D.}\ \bibnamefont {O'Connell}}, \bibinfo {author}
  {\bibfnamefont {D.}~\bibnamefont {Sank}}, \bibinfo {author} {\bibfnamefont
  {H.}~\bibnamefont {Wang}}, \bibinfo {author} {\bibfnamefont {M.}~\bibnamefont
  {Weides}}, \bibinfo {author} {\bibfnamefont {J.}~\bibnamefont {Wenner}},
  \bibinfo {author} {\bibfnamefont {Y.}~\bibnamefont {Yin}}, \bibinfo {author}
  {\bibfnamefont {T.}~\bibnamefont {Yamamoto}}, \bibinfo {author}
  {\bibfnamefont {A.~N.}\ \bibnamefont {Cleland}}, \ and\ \bibinfo {author}
  {\bibfnamefont {J.~M.}\ \bibnamefont {Martinis}},\ }\href {\doibase
  10.1038/nature09418} {\bibfield  {journal} {\bibinfo  {journal} {Nature}\
  }\textbf {\bibinfo {volume} {467}},\ \bibinfo {pages} {570} (\bibinfo {year}
  {2010})}\BibitemShut {NoStop}%
\bibitem [{\citenamefont {Mosley}\ \emph {et~al.}(2008)\citenamefont {Mosley},
  \citenamefont {Lundeen}, \citenamefont {Smith}, \citenamefont {Wasylczyk},
  \citenamefont {U'Ren}, \citenamefont {Silberhorn},\ and\ \citenamefont
  {Walmsley}}]{Mosley2008}%
  \BibitemOpen
  \bibfield  {author} {\bibinfo {author} {\bibfnamefont {P.~J.}\ \bibnamefont
  {Mosley}}, \bibinfo {author} {\bibfnamefont {J.~S.}\ \bibnamefont {Lundeen}},
  \bibinfo {author} {\bibfnamefont {B.~J.}\ \bibnamefont {Smith}}, \bibinfo
  {author} {\bibfnamefont {P.}~\bibnamefont {Wasylczyk}}, \bibinfo {author}
  {\bibfnamefont {A.~B.}\ \bibnamefont {U'Ren}}, \bibinfo {author}
  {\bibfnamefont {C.}~\bibnamefont {Silberhorn}}, \ and\ \bibinfo {author}
  {\bibfnamefont {I.~A.}\ \bibnamefont {Walmsley}},\ }\href {\doibase
  10.1103/PhysRevLett.100.133601} {\bibfield  {journal} {\bibinfo  {journal}
  {Physical Review Letters}\ }\textbf {\bibinfo {volume} {100}},\ \bibinfo
  {pages} {133601} (\bibinfo {year} {2008})}\BibitemShut {NoStop}%
\bibitem [{\citenamefont {Azuma}\ \emph {et~al.}(2015)\citenamefont {Azuma},
  \citenamefont {Tamaki},\ and\ \citenamefont {Lo}}]{Azuma2015}%
  \BibitemOpen
  \bibfield  {author} {\bibinfo {author} {\bibfnamefont {K.}~\bibnamefont
  {Azuma}}, \bibinfo {author} {\bibfnamefont {K.}~\bibnamefont {Tamaki}}, \
  and\ \bibinfo {author} {\bibfnamefont {H.-K.}\ \bibnamefont {Lo}},\ }\href
  {\doibase 10.1038/ncomms7787} {\bibfield  {journal} {\bibinfo  {journal}
  {Nature Communications}\ }\textbf {\bibinfo {volume} {6}},\ \bibinfo {pages}
  {6787} (\bibinfo {year} {2015})}\BibitemShut {NoStop}%
\bibitem [{\citenamefont {Kurpiers}\ \emph {et~al.}(2019)\citenamefont
  {Kurpiers}, \citenamefont {Pechal}, \citenamefont {Royer}, \citenamefont
  {Magnard}, \citenamefont {Walter}, \citenamefont {Heinsoo}, \citenamefont
  {Salathe}, \citenamefont {Akin}, \citenamefont {Storz}, \citenamefont
  {Besse}, \citenamefont {Gasparinetti}, \citenamefont {Blais},\ and\
  \citenamefont {Wallraff}}]{Kurpiers2019}%
  \BibitemOpen
  \bibfield  {author} {\bibinfo {author} {\bibfnamefont {P.}~\bibnamefont
  {Kurpiers}}, \bibinfo {author} {\bibfnamefont {M.}~\bibnamefont {Pechal}},
  \bibinfo {author} {\bibfnamefont {B.}~\bibnamefont {Royer}}, \bibinfo
  {author} {\bibfnamefont {P.}~\bibnamefont {Magnard}}, \bibinfo {author}
  {\bibfnamefont {T.}~\bibnamefont {Walter}}, \bibinfo {author} {\bibfnamefont
  {J.}~\bibnamefont {Heinsoo}}, \bibinfo {author} {\bibfnamefont
  {Y.}~\bibnamefont {Salathe}}, \bibinfo {author} {\bibfnamefont
  {A.}~\bibnamefont {Akin}}, \bibinfo {author} {\bibfnamefont {S.}~\bibnamefont
  {Storz}}, \bibinfo {author} {\bibfnamefont {J.-C.}\ \bibnamefont {Besse}},
  \bibinfo {author} {\bibfnamefont {S.}~\bibnamefont {Gasparinetti}}, \bibinfo
  {author} {\bibfnamefont {A.}~\bibnamefont {Blais}}, \ and\ \bibinfo {author}
  {\bibfnamefont {A.}~\bibnamefont {Wallraff}},\ }\href {\doibase
  10.1103/PhysRevApplied.12.044067} {\bibfield  {journal} {\bibinfo  {journal}
  {Physical Review Applied}\ }\textbf {\bibinfo {volume} {12}},\ \bibinfo
  {pages} {044067} (\bibinfo {year} {2019})}\BibitemShut {NoStop}%
\bibitem [{\citenamefont {Kwiat}\ \emph {et~al.}(2001)\citenamefont {Kwiat},
  \citenamefont {Barraza-Lopez}, \citenamefont {Stefanov},\ and\ \citenamefont
  {Gisin}}]{Kwiat2001}%
  \BibitemOpen
  \bibfield  {author} {\bibinfo {author} {\bibfnamefont {P.~G.}\ \bibnamefont
  {Kwiat}}, \bibinfo {author} {\bibfnamefont {S.}~\bibnamefont
  {Barraza-Lopez}}, \bibinfo {author} {\bibfnamefont {A.}~\bibnamefont
  {Stefanov}}, \ and\ \bibinfo {author} {\bibfnamefont {N.}~\bibnamefont
  {Gisin}},\ }\href {\doibase 10.1038/35059017} {\bibfield  {journal} {\bibinfo
   {journal} {Nature}\ }\textbf {\bibinfo {volume} {409}},\ \bibinfo {pages}
  {1014} (\bibinfo {year} {2001})}\BibitemShut {NoStop}%
\bibitem [{\citenamefont {Dong}\ \emph {et~al.}(2008)\citenamefont {Dong},
  \citenamefont {Lassen}, \citenamefont {Heersink}, \citenamefont {Marquardt},
  \citenamefont {Filip}, \citenamefont {Leuchs},\ and\ \citenamefont
  {Andersen}}]{Dong2008}%
  \BibitemOpen
  \bibfield  {author} {\bibinfo {author} {\bibfnamefont {R.}~\bibnamefont
  {Dong}}, \bibinfo {author} {\bibfnamefont {M.}~\bibnamefont {Lassen}},
  \bibinfo {author} {\bibfnamefont {J.}~\bibnamefont {Heersink}}, \bibinfo
  {author} {\bibfnamefont {C.}~\bibnamefont {Marquardt}}, \bibinfo {author}
  {\bibfnamefont {R.}~\bibnamefont {Filip}}, \bibinfo {author} {\bibfnamefont
  {G.}~\bibnamefont {Leuchs}}, \ and\ \bibinfo {author} {\bibfnamefont {U.~L.}\
  \bibnamefont {Andersen}},\ }\href {\doibase 10.1038/nphys1112} {\bibfield
  {journal} {\bibinfo  {journal} {Nature Physics}\ }\textbf {\bibinfo {volume}
  {4}},\ \bibinfo {pages} {919} (\bibinfo {year} {2008})}\BibitemShut {NoStop}%
\bibitem [{\citenamefont {Takahashi}\ \emph {et~al.}(2010)\citenamefont
  {Takahashi}, \citenamefont {Neergaard-Nielsen}, \citenamefont {Takeuchi},
  \citenamefont {Takeoka}, \citenamefont {Hayasaka}, \citenamefont {Furusawa},\
  and\ \citenamefont {Sasaki}}]{Takahashi2010}%
  \BibitemOpen
  \bibfield  {author} {\bibinfo {author} {\bibfnamefont {H.}~\bibnamefont
  {Takahashi}}, \bibinfo {author} {\bibfnamefont {J.~S.}\ \bibnamefont
  {Neergaard-Nielsen}}, \bibinfo {author} {\bibfnamefont {M.}~\bibnamefont
  {Takeuchi}}, \bibinfo {author} {\bibfnamefont {M.}~\bibnamefont {Takeoka}},
  \bibinfo {author} {\bibfnamefont {K.}~\bibnamefont {Hayasaka}}, \bibinfo
  {author} {\bibfnamefont {A.}~\bibnamefont {Furusawa}}, \ and\ \bibinfo
  {author} {\bibfnamefont {M.}~\bibnamefont {Sasaki}},\ }\href {\doibase
  10.1038/nphoton.2010.1} {\bibfield  {journal} {\bibinfo  {journal} {Nature
  Photonics}\ }\textbf {\bibinfo {volume} {4}},\ \bibinfo {pages} {178}
  (\bibinfo {year} {2010})}\BibitemShut {NoStop}%
\bibitem [{\citenamefont {Poyatos}\ \emph {et~al.}(1996)\citenamefont
  {Poyatos}, \citenamefont {Cirac},\ and\ \citenamefont
  {Zoller}}]{Poyatos1996}%
  \BibitemOpen
  \bibfield  {author} {\bibinfo {author} {\bibfnamefont {J.~F.}\ \bibnamefont
  {Poyatos}}, \bibinfo {author} {\bibfnamefont {J.~I.}\ \bibnamefont {Cirac}},
  \ and\ \bibinfo {author} {\bibfnamefont {P.}~\bibnamefont {Zoller}},\ }\href
  {\doibase 10.1103/PhysRevLett.77.4728} {\bibfield  {journal} {\bibinfo
  {journal} {Physical Review Letters}\ }\textbf {\bibinfo {volume} {77}},\
  \bibinfo {pages} {4728} (\bibinfo {year} {1996})}\BibitemShut {NoStop}%
\bibitem [{\citenamefont {Plenio}\ and\ \citenamefont
  {Huelga}(2002)}]{Plenio2002}%
  \BibitemOpen
  \bibfield  {author} {\bibinfo {author} {\bibfnamefont {M.~B.}\ \bibnamefont
  {Plenio}}\ and\ \bibinfo {author} {\bibfnamefont {S.~F.}\ \bibnamefont
  {Huelga}},\ }\href {\doibase 10.1103/PhysRevLett.88.197901} {\bibfield
  {journal} {\bibinfo  {journal} {Physical Review Letters}\ }\textbf {\bibinfo
  {volume} {88}},\ \bibinfo {pages} {197901} (\bibinfo {year}
  {2002})}\BibitemShut {NoStop}%
\bibitem [{\citenamefont {Hermelin}\ \emph {et~al.}(2011)\citenamefont
  {Hermelin}, \citenamefont {Takada}, \citenamefont {Yamamoto}, \citenamefont
  {Tarucha}, \citenamefont {Wieck}, \citenamefont {Saminadayar}, \citenamefont
  {Bauerle},\ and\ \citenamefont {Meunier}}]{Hermelin2011}%
  \BibitemOpen
  \bibfield  {author} {\bibinfo {author} {\bibfnamefont {S.}~\bibnamefont
  {Hermelin}}, \bibinfo {author} {\bibfnamefont {S.}~\bibnamefont {Takada}},
  \bibinfo {author} {\bibfnamefont {M.}~\bibnamefont {Yamamoto}}, \bibinfo
  {author} {\bibfnamefont {S.}~\bibnamefont {Tarucha}}, \bibinfo {author}
  {\bibfnamefont {A.~D.}\ \bibnamefont {Wieck}}, \bibinfo {author}
  {\bibfnamefont {L.}~\bibnamefont {Saminadayar}}, \bibinfo {author}
  {\bibfnamefont {C.}~\bibnamefont {Bauerle}}, \ and\ \bibinfo {author}
  {\bibfnamefont {T.}~\bibnamefont {Meunier}},\ }\href {\doibase
  10.1038/nature10416} {\bibfield  {journal} {\bibinfo  {journal} {Nature}\
  }\textbf {\bibinfo {volume} {477}},\ \bibinfo {pages} {435} (\bibinfo {year}
  {2011})}\BibitemShut {NoStop}%
\bibitem [{\citenamefont {McNeil}\ \emph {et~al.}(2011)\citenamefont {McNeil},
  \citenamefont {Kataoka}, \citenamefont {Ford}, \citenamefont {Barnes},
  \citenamefont {Anderson}, \citenamefont {Jones}, \citenamefont {Farrer},\
  and\ \citenamefont {Ritchie}}]{McNeil2011}%
  \BibitemOpen
  \bibfield  {author} {\bibinfo {author} {\bibfnamefont {R.~P.~G.}\
  \bibnamefont {McNeil}}, \bibinfo {author} {\bibfnamefont {M.}~\bibnamefont
  {Kataoka}}, \bibinfo {author} {\bibfnamefont {C.~J.~B.}\ \bibnamefont
  {Ford}}, \bibinfo {author} {\bibfnamefont {C.~H.~W.}\ \bibnamefont {Barnes}},
  \bibinfo {author} {\bibfnamefont {D.}~\bibnamefont {Anderson}}, \bibinfo
  {author} {\bibfnamefont {G.~A.~C.}\ \bibnamefont {Jones}}, \bibinfo {author}
  {\bibfnamefont {I.}~\bibnamefont {Farrer}}, \ and\ \bibinfo {author}
  {\bibfnamefont {D.~A.}\ \bibnamefont {Ritchie}},\ }\href {\doibase
  10.1038/nature10444} {\bibfield  {journal} {\bibinfo  {journal} {Nature}\
  }\textbf {\bibinfo {volume} {477}},\ \bibinfo {pages} {439} (\bibinfo {year}
  {2011})}\BibitemShut {NoStop}%
\bibitem [{\citenamefont {Baksic}\ \emph {et~al.}(2016)\citenamefont {Baksic},
  \citenamefont {Ribeiro},\ and\ \citenamefont {Clerk}}]{Baksic2016}%
  \BibitemOpen
  \bibfield  {author} {\bibinfo {author} {\bibfnamefont {A.}~\bibnamefont
  {Baksic}}, \bibinfo {author} {\bibfnamefont {H.}~\bibnamefont {Ribeiro}}, \
  and\ \bibinfo {author} {\bibfnamefont {A.~A.}\ \bibnamefont {Clerk}},\ }\href
  {\doibase 10.1103/PhysRevLett.116.230503} {\bibfield  {journal} {\bibinfo
  {journal} {Physical Review Letters}\ }\textbf {\bibinfo {volume} {116}},\
  \bibinfo {pages} {230503} (\bibinfo {year} {2016})}\BibitemShut {NoStop}%
\bibitem [{\citenamefont {Zhou}\ \emph {et~al.}(2017)\citenamefont {Zhou},
  \citenamefont {Baksic}, \citenamefont {Ribeiro}, \citenamefont {Yale},
  \citenamefont {Heremans}, \citenamefont {Jerger}, \citenamefont {Auer},
  \citenamefont {Burkard}, \citenamefont {Clerk},\ and\ \citenamefont
  {Awschalom}}]{Zhou2017}%
  \BibitemOpen
  \bibfield  {author} {\bibinfo {author} {\bibfnamefont {B.~B.}\ \bibnamefont
  {Zhou}}, \bibinfo {author} {\bibfnamefont {A.}~\bibnamefont {Baksic}},
  \bibinfo {author} {\bibfnamefont {H.}~\bibnamefont {Ribeiro}}, \bibinfo
  {author} {\bibfnamefont {C.~G.}\ \bibnamefont {Yale}}, \bibinfo {author}
  {\bibfnamefont {F.~J.}\ \bibnamefont {Heremans}}, \bibinfo {author}
  {\bibfnamefont {P.~C.}\ \bibnamefont {Jerger}}, \bibinfo {author}
  {\bibfnamefont {A.}~\bibnamefont {Auer}}, \bibinfo {author} {\bibfnamefont
  {G.}~\bibnamefont {Burkard}}, \bibinfo {author} {\bibfnamefont {A.~A.}\
  \bibnamefont {Clerk}}, \ and\ \bibinfo {author} {\bibfnamefont {D.~D.}\
  \bibnamefont {Awschalom}},\ }\href {\doibase 10.1038/nphys3967} {\bibfield
  {journal} {\bibinfo  {journal} {Nature Physics}\ }\textbf {\bibinfo {volume}
  {13}},\ \bibinfo {pages} {330} (\bibinfo {year} {2017})}\BibitemShut
  {NoStop}%
\bibitem [{\citenamefont {Torosov}\ \emph {et~al.}(2011)\citenamefont
  {Torosov}, \citenamefont {Guerin},\ and\ \citenamefont
  {Vitanov}}]{Torosov2011}%
  \BibitemOpen
  \bibfield  {author} {\bibinfo {author} {\bibfnamefont {B.~T.}\ \bibnamefont
  {Torosov}}, \bibinfo {author} {\bibfnamefont {S.}~\bibnamefont {Guerin}}, \
  and\ \bibinfo {author} {\bibfnamefont {N.~V.}\ \bibnamefont {Vitanov}},\
  }\href {\doibase 10.1103/PhysRevLett.106.233001} {\bibfield  {journal}
  {\bibinfo  {journal} {Physical Review Letters}\ }\textbf {\bibinfo {volume}
  {106}},\ \bibinfo {pages} {233001} (\bibinfo {year} {2011})}\BibitemShut
  {NoStop}%
\bibitem [{\citenamefont {Bruns}\ \emph {et~al.}(2018)\citenamefont {Bruns},
  \citenamefont {Genov}, \citenamefont {Hain}, \citenamefont {Vitanov},\ and\
  \citenamefont {Halfmann}}]{Bruns2018}%
  \BibitemOpen
  \bibfield  {author} {\bibinfo {author} {\bibfnamefont {A.}~\bibnamefont
  {Bruns}}, \bibinfo {author} {\bibfnamefont {G.~T.}\ \bibnamefont {Genov}},
  \bibinfo {author} {\bibfnamefont {M.}~\bibnamefont {Hain}}, \bibinfo {author}
  {\bibfnamefont {N.~V.}\ \bibnamefont {Vitanov}}, \ and\ \bibinfo {author}
  {\bibfnamefont {T.}~\bibnamefont {Halfmann}},\ }\href {\doibase
  10.1103/PhysRevA.98.053413} {\bibfield  {journal} {\bibinfo  {journal}
  {Physical Review A}\ }\textbf {\bibinfo {volume} {98}},\ \bibinfo {pages}
  {053413} (\bibinfo {year} {2018})}\BibitemShut {NoStop}%
\end{thebibliography}%


\begin{thebibliography}{28}%
\makeatletter
\providecommand \@ifxundefined [1]{%
 \@ifx{#1\undefined}
}%
\providecommand \@ifnum [1]{%
 \ifnum #1\expandafter \@firstoftwo
 \else \expandafter \@secondoftwo
 \fi
}%
\providecommand \@ifx [1]{%
 \ifx #1\expandafter \@firstoftwo
 \else \expandafter \@secondoftwo
 \fi
}%
\providecommand \natexlab [1]{#1}%
\providecommand \enquote  [1]{``#1''}%
\providecommand \bibnamefont  [1]{#1}%
\providecommand \bibfnamefont [1]{#1}%
\providecommand \citenamefont [1]{#1}%
\providecommand \href@noop [0]{\@secondoftwo}%
\providecommand \href [0]{\begingroup \@sanitize@url \@href}%
\providecommand \@href[1]{\@@startlink{#1}\@@href}%
\providecommand \@@href[1]{\endgroup#1\@@endlink}%
\providecommand \@sanitize@url [0]{\catcode `\\12\catcode `\$12\catcode
  `\&12\catcode `\#12\catcode `\^12\catcode `\_12\catcode `\%12\relax}%
\providecommand \@@startlink[1]{}%
\providecommand \@@endlink[0]{}%
\providecommand \url  [0]{\begingroup\@sanitize@url \@url }%
\providecommand \@url [1]{\endgroup\@href {#1}{\urlprefix }}%
\providecommand \urlprefix  [0]{URL }%
\providecommand \Eprint [0]{\href }%
\providecommand \doibase [0]{http://dx.doi.org/}%
\providecommand \selectlanguage [0]{\@gobble}%
\providecommand \bibinfo  [0]{\@secondoftwo}%
\providecommand \bibfield  [0]{\@secondoftwo}%
\providecommand \translation [1]{[#1]}%
\providecommand \BibitemOpen [0]{}%
\providecommand \bibitemStop [0]{}%
\providecommand \bibitemNoStop [0]{.\EOS\space}%
\providecommand \EOS [0]{\spacefactor3000\relax}%
\providecommand \BibitemShut  [1]{\csname bibitem#1\endcsname}%
\let\auto@bib@innerbib\@empty
%</preamble>
\bibitem [{\citenamefont {Zhong}\ \emph {et~al.}(2019)\citenamefont {Zhong},
  \citenamefont {Chang}, \citenamefont {Satzinger}, \citenamefont {Chou},
  \citenamefont {Bienfait}, \citenamefont {Conner}, \citenamefont {Dumur},
  \citenamefont {Grebel}, \citenamefont {Peairs}, \citenamefont {Povey},
  \citenamefont {Schuster},\ and\ \citenamefont {Cleland}}]{Zhong2019}%
  \BibitemOpen
  \bibfield  {author} {\bibinfo {author} {\bibfnamefont {Y.~P.}\ \bibnamefont
  {Zhong}}, \bibinfo {author} {\bibfnamefont {H.-S.}\ \bibnamefont {Chang}},
  \bibinfo {author} {\bibfnamefont {K.~J.}\ \bibnamefont {Satzinger}}, \bibinfo
  {author} {\bibfnamefont {M.-H.}\ \bibnamefont {Chou}}, \bibinfo {author}
  {\bibfnamefont {A.}~\bibnamefont {Bienfait}}, \bibinfo {author}
  {\bibfnamefont {C.~R.}\ \bibnamefont {Conner}}, \bibinfo {author}
  {\bibfnamefont {{\'E}.}~\bibnamefont {Dumur}}, \bibinfo {author}
  {\bibfnamefont {J.}~\bibnamefont {Grebel}}, \bibinfo {author} {\bibfnamefont
  {G.~A.}\ \bibnamefont {Peairs}}, \bibinfo {author} {\bibfnamefont {R.~G.}\
  \bibnamefont {Povey}}, \bibinfo {author} {\bibfnamefont {D.~I.}\ \bibnamefont
  {Schuster}}, \ and\ \bibinfo {author} {\bibfnamefont {A.~N.}\ \bibnamefont
  {Cleland}},\ }\href {\doibase 10.1038/s41567-019-0507-7} {\bibfield
  {journal} {\bibinfo  {journal} {Nature Physics}\ }\textbf {\bibinfo {volume}
  {15}},\ \bibinfo {pages} {741} (\bibinfo {year} {2019})}\BibitemShut
  {NoStop}%
\bibitem [{\citenamefont {Koch}\ \emph {et~al.}(2007)\citenamefont {Koch},
  \citenamefont {Yu}, \citenamefont {Gambetta}, \citenamefont {Houck},
  \citenamefont {Schuster}, \citenamefont {Majer}, \citenamefont {Blais},
  \citenamefont {Devoret}, \citenamefont {Girvin},\ and\ \citenamefont
  {Schoelkopf}}]{Koch2007}%
  \BibitemOpen
  \bibfield  {author} {\bibinfo {author} {\bibfnamefont {J.}~\bibnamefont
  {Koch}}, \bibinfo {author} {\bibfnamefont {T.~M.}\ \bibnamefont {Yu}},
  \bibinfo {author} {\bibfnamefont {J.}~\bibnamefont {Gambetta}}, \bibinfo
  {author} {\bibfnamefont {A.~A.}\ \bibnamefont {Houck}}, \bibinfo {author}
  {\bibfnamefont {D.~I.}\ \bibnamefont {Schuster}}, \bibinfo {author}
  {\bibfnamefont {J.}~\bibnamefont {Majer}}, \bibinfo {author} {\bibfnamefont
  {A.}~\bibnamefont {Blais}}, \bibinfo {author} {\bibfnamefont {M.~H.}\
  \bibnamefont {Devoret}}, \bibinfo {author} {\bibfnamefont {S.~M.}\
  \bibnamefont {Girvin}}, \ and\ \bibinfo {author} {\bibfnamefont {R.~J.}\
  \bibnamefont {Schoelkopf}},\ }\href {\doibase 10.1103/PhysRevA.76.042319}
  {\bibfield  {journal} {\bibinfo  {journal} {Physical Review A}\ }\textbf
  {\bibinfo {volume} {76}},\ \bibinfo {pages} {042319} (\bibinfo {year}
  {2007})}\BibitemShut {NoStop}%
\bibitem [{\citenamefont {Barends}\ \emph {et~al.}(2013)\citenamefont
  {Barends}, \citenamefont {Kelly}, \citenamefont {Megrant}, \citenamefont
  {Sank}, \citenamefont {Jeffrey}, \citenamefont {Chen}, \citenamefont {Yin},
  \citenamefont {Chiaro}, \citenamefont {Mutus}, \citenamefont {Neill},
  \citenamefont {O'Malley}, \citenamefont {Roushan}, \citenamefont {Wenner},
  \citenamefont {White}, \citenamefont {Cleland},\ and\ \citenamefont
  {Martinis}}]{Barends2013}%
  \BibitemOpen
  \bibfield  {author} {\bibinfo {author} {\bibfnamefont {R.}~\bibnamefont
  {Barends}}, \bibinfo {author} {\bibfnamefont {J.}~\bibnamefont {Kelly}},
  \bibinfo {author} {\bibfnamefont {A.}~\bibnamefont {Megrant}}, \bibinfo
  {author} {\bibfnamefont {D.}~\bibnamefont {Sank}}, \bibinfo {author}
  {\bibfnamefont {E.}~\bibnamefont {Jeffrey}}, \bibinfo {author} {\bibfnamefont
  {Y.}~\bibnamefont {Chen}}, \bibinfo {author} {\bibfnamefont {Y.}~\bibnamefont
  {Yin}}, \bibinfo {author} {\bibfnamefont {B.}~\bibnamefont {Chiaro}},
  \bibinfo {author} {\bibfnamefont {J.}~\bibnamefont {Mutus}}, \bibinfo
  {author} {\bibfnamefont {C.}~\bibnamefont {Neill}}, \bibinfo {author}
  {\bibfnamefont {P.}~\bibnamefont {O'Malley}}, \bibinfo {author}
  {\bibfnamefont {P.}~\bibnamefont {Roushan}}, \bibinfo {author} {\bibfnamefont
  {J.}~\bibnamefont {Wenner}}, \bibinfo {author} {\bibfnamefont {T.~C.}\
  \bibnamefont {White}}, \bibinfo {author} {\bibfnamefont {A.~N.}\ \bibnamefont
  {Cleland}}, \ and\ \bibinfo {author} {\bibfnamefont {J.~M.}\ \bibnamefont
  {Martinis}},\ }\href {\doibase 10.1103/PhysRevLett.111.080502} {\bibfield
  {journal} {\bibinfo  {journal} {Physical Review Letters}\ }\textbf {\bibinfo
  {volume} {111}},\ \bibinfo {pages} {080502} (\bibinfo {year}
  {2013})}\BibitemShut {NoStop}%
\bibitem [{\citenamefont {Jeffrey}\ \emph {et~al.}(2014)\citenamefont
  {Jeffrey}, \citenamefont {Sank}, \citenamefont {Mutus}, \citenamefont
  {White}, \citenamefont {Kelly}, \citenamefont {Barends}, \citenamefont
  {Chen}, \citenamefont {Chen}, \citenamefont {Chiaro}, \citenamefont
  {Dunsworth}, \citenamefont {Megrant}, \citenamefont {O'Malley}, \citenamefont
  {Neill}, \citenamefont {Roushan}, \citenamefont {Vainsencher}, \citenamefont
  {Wenner}, \citenamefont {Cleland},\ and\ \citenamefont
  {Martinis}}]{Jeffrey2014}%
  \BibitemOpen
  \bibfield  {author} {\bibinfo {author} {\bibfnamefont {E.}~\bibnamefont
  {Jeffrey}}, \bibinfo {author} {\bibfnamefont {D.}~\bibnamefont {Sank}},
  \bibinfo {author} {\bibfnamefont {J.~Y.}\ \bibnamefont {Mutus}}, \bibinfo
  {author} {\bibfnamefont {T.~C.}\ \bibnamefont {White}}, \bibinfo {author}
  {\bibfnamefont {J.}~\bibnamefont {Kelly}}, \bibinfo {author} {\bibfnamefont
  {R.}~\bibnamefont {Barends}}, \bibinfo {author} {\bibfnamefont
  {Y.}~\bibnamefont {Chen}}, \bibinfo {author} {\bibfnamefont {Z.}~\bibnamefont
  {Chen}}, \bibinfo {author} {\bibfnamefont {B.}~\bibnamefont {Chiaro}},
  \bibinfo {author} {\bibfnamefont {A.}~\bibnamefont {Dunsworth}}, \bibinfo
  {author} {\bibfnamefont {A.}~\bibnamefont {Megrant}}, \bibinfo {author}
  {\bibfnamefont {P.~J.~J.}\ \bibnamefont {O'Malley}}, \bibinfo {author}
  {\bibfnamefont {C.}~\bibnamefont {Neill}}, \bibinfo {author} {\bibfnamefont
  {P.}~\bibnamefont {Roushan}}, \bibinfo {author} {\bibfnamefont
  {A.}~\bibnamefont {Vainsencher}}, \bibinfo {author} {\bibfnamefont
  {J.}~\bibnamefont {Wenner}}, \bibinfo {author} {\bibfnamefont {A.~N.}\
  \bibnamefont {Cleland}}, \ and\ \bibinfo {author} {\bibfnamefont {J.~M.}\
  \bibnamefont {Martinis}},\ }\href {\doibase 10.1103/PhysRevLett.112.190504}
  {\bibfield  {journal} {\bibinfo  {journal} {Physical Review Letters}\
  }\textbf {\bibinfo {volume} {112}},\ \bibinfo {pages} {190504} (\bibinfo
  {year} {2014})}\BibitemShut {NoStop}%
\bibitem [{\citenamefont {Kelly}\ \emph {et~al.}(2015)\citenamefont {Kelly},
  \citenamefont {Barends}, \citenamefont {Fowler}, \citenamefont {Megrant},
  \citenamefont {Jeffrey}, \citenamefont {White}, \citenamefont {Sank},
  \citenamefont {Mutus}, \citenamefont {Campbell}, \citenamefont {Chen},
  \citenamefont {Chen}, \citenamefont {Chiaro}, \citenamefont {Dunsworth},
  \citenamefont {Hoi}, \citenamefont {Neill}, \citenamefont {O'Malley},
  \citenamefont {Quintana}, \citenamefont {Roushan}, \citenamefont
  {Vainsencher}, \citenamefont {Wenner}, \citenamefont {Cleland},\ and\
  \citenamefont {Martinis}}]{Kelly2015}%
  \BibitemOpen
  \bibfield  {author} {\bibinfo {author} {\bibfnamefont {J.}~\bibnamefont
  {Kelly}}, \bibinfo {author} {\bibfnamefont {R.}~\bibnamefont {Barends}},
  \bibinfo {author} {\bibfnamefont {A.~G.}\ \bibnamefont {Fowler}}, \bibinfo
  {author} {\bibfnamefont {A.}~\bibnamefont {Megrant}}, \bibinfo {author}
  {\bibfnamefont {E.}~\bibnamefont {Jeffrey}}, \bibinfo {author} {\bibfnamefont
  {T.~C.}\ \bibnamefont {White}}, \bibinfo {author} {\bibfnamefont
  {D.}~\bibnamefont {Sank}}, \bibinfo {author} {\bibfnamefont {J.~Y.}\
  \bibnamefont {Mutus}}, \bibinfo {author} {\bibfnamefont {B.}~\bibnamefont
  {Campbell}}, \bibinfo {author} {\bibfnamefont {Y.}~\bibnamefont {Chen}},
  \bibinfo {author} {\bibfnamefont {Z.}~\bibnamefont {Chen}}, \bibinfo {author}
  {\bibfnamefont {B.}~\bibnamefont {Chiaro}}, \bibinfo {author} {\bibfnamefont
  {A.}~\bibnamefont {Dunsworth}}, \bibinfo {author} {\bibfnamefont {I.-C.}\
  \bibnamefont {Hoi}}, \bibinfo {author} {\bibfnamefont {C.}~\bibnamefont
  {Neill}}, \bibinfo {author} {\bibfnamefont {P.~J.~J.}\ \bibnamefont
  {O'Malley}}, \bibinfo {author} {\bibfnamefont {C.}~\bibnamefont {Quintana}},
  \bibinfo {author} {\bibfnamefont {P.}~\bibnamefont {Roushan}}, \bibinfo
  {author} {\bibfnamefont {A.}~\bibnamefont {Vainsencher}}, \bibinfo {author}
  {\bibfnamefont {J.}~\bibnamefont {Wenner}}, \bibinfo {author} {\bibfnamefont
  {A.~N.}\ \bibnamefont {Cleland}}, \ and\ \bibinfo {author} {\bibfnamefont
  {J.~M.}\ \bibnamefont {Martinis}},\ }\href {\doibase 10.1038/nature14270}
  {\bibfield  {journal} {\bibinfo  {journal} {Nature}\ }\textbf {\bibinfo
  {volume} {519}},\ \bibinfo {pages} {66} (\bibinfo {year} {2015})}\BibitemShut
  {NoStop}%
\bibitem [{\citenamefont {Macklin}\ \emph {et~al.}(2015)\citenamefont
  {Macklin}, \citenamefont {O'Brien}, \citenamefont {Hover}, \citenamefont
  {Schwartz}, \citenamefont {Bolkhovsky}, \citenamefont {Zhang}, \citenamefont
  {Oliver},\ and\ \citenamefont {Siddiqi}}]{Macklin2015}%
  \BibitemOpen
  \bibfield  {author} {\bibinfo {author} {\bibfnamefont {C.}~\bibnamefont
  {Macklin}}, \bibinfo {author} {\bibfnamefont {K.}~\bibnamefont {O'Brien}},
  \bibinfo {author} {\bibfnamefont {D.}~\bibnamefont {Hover}}, \bibinfo
  {author} {\bibfnamefont {M.~E.}\ \bibnamefont {Schwartz}}, \bibinfo {author}
  {\bibfnamefont {V.}~\bibnamefont {Bolkhovsky}}, \bibinfo {author}
  {\bibfnamefont {X.}~\bibnamefont {Zhang}}, \bibinfo {author} {\bibfnamefont
  {W.~D.}\ \bibnamefont {Oliver}}, \ and\ \bibinfo {author} {\bibfnamefont
  {I.}~\bibnamefont {Siddiqi}},\ }\href {\doibase 10.1126/science.aaa8525}
  {\bibfield  {journal} {\bibinfo  {journal} {Science}\ }\textbf {\bibinfo
  {volume} {350}},\ \bibinfo {pages} {307} (\bibinfo {year}
  {2015})}\BibitemShut {NoStop}%
\bibitem [{\citenamefont {Chen}\ \emph {et~al.}(2014)\citenamefont {Chen},
  \citenamefont {Neill}, \citenamefont {Roushan}, \citenamefont {Leung},
  \citenamefont {Fang}, \citenamefont {Barends}, \citenamefont {Kelly},
  \citenamefont {Campbell}, \citenamefont {Chen}, \citenamefont {Chiaro},
  \citenamefont {Dunsworth}, \citenamefont {Jeffrey}, \citenamefont {Megrant},
  \citenamefont {Mutus}, \citenamefont {O'Malley}, \citenamefont {Quintana},
  \citenamefont {Sank}, \citenamefont {Vainsencher}, \citenamefont {Wenner},
  \citenamefont {White}, \citenamefont {Geller}, \citenamefont {Cleland},\ and\
  \citenamefont {Martinis}}]{Chen2014}%
  \BibitemOpen
  \bibfield  {author} {\bibinfo {author} {\bibfnamefont {Y.}~\bibnamefont
  {Chen}}, \bibinfo {author} {\bibfnamefont {C.}~\bibnamefont {Neill}},
  \bibinfo {author} {\bibfnamefont {P.}~\bibnamefont {Roushan}}, \bibinfo
  {author} {\bibfnamefont {N.}~\bibnamefont {Leung}}, \bibinfo {author}
  {\bibfnamefont {M.}~\bibnamefont {Fang}}, \bibinfo {author} {\bibfnamefont
  {R.}~\bibnamefont {Barends}}, \bibinfo {author} {\bibfnamefont
  {J.}~\bibnamefont {Kelly}}, \bibinfo {author} {\bibfnamefont
  {B.}~\bibnamefont {Campbell}}, \bibinfo {author} {\bibfnamefont
  {Z.}~\bibnamefont {Chen}}, \bibinfo {author} {\bibfnamefont {B.}~\bibnamefont
  {Chiaro}}, \bibinfo {author} {\bibfnamefont {A.}~\bibnamefont {Dunsworth}},
  \bibinfo {author} {\bibfnamefont {E.}~\bibnamefont {Jeffrey}}, \bibinfo
  {author} {\bibfnamefont {A.}~\bibnamefont {Megrant}}, \bibinfo {author}
  {\bibfnamefont {J.~Y.}\ \bibnamefont {Mutus}}, \bibinfo {author}
  {\bibfnamefont {P.~J.~J.}\ \bibnamefont {O'Malley}}, \bibinfo {author}
  {\bibfnamefont {C.~M.}\ \bibnamefont {Quintana}}, \bibinfo {author}
  {\bibfnamefont {D.}~\bibnamefont {Sank}}, \bibinfo {author} {\bibfnamefont
  {A.}~\bibnamefont {Vainsencher}}, \bibinfo {author} {\bibfnamefont
  {J.}~\bibnamefont {Wenner}}, \bibinfo {author} {\bibfnamefont {T.~C.}\
  \bibnamefont {White}}, \bibinfo {author} {\bibfnamefont {M.~R.}\ \bibnamefont
  {Geller}}, \bibinfo {author} {\bibfnamefont {A.~N.}\ \bibnamefont {Cleland}},
  \ and\ \bibinfo {author} {\bibfnamefont {J.~M.}\ \bibnamefont {Martinis}},\
  }\href {\doibase 10.1103/PhysRevLett.113.220502} {\bibfield  {journal}
  {\bibinfo  {journal} {Physical Review Letters}\ }\textbf {\bibinfo {volume}
  {113}},\ \bibinfo {pages} {220502} (\bibinfo {year} {2014})}\BibitemShut
  {NoStop}%
\bibitem [{\citenamefont {Pozar}(2012)}]{Pozar}%
  \BibitemOpen
  \bibfield  {author} {\bibinfo {author} {\bibfnamefont {D.~M.}\ \bibnamefont
  {Pozar}},\ }\href@noop {} {\emph {\bibinfo {title} {Microwave
  engineering}}},\ \bibinfo {edition} {4th}\ ed.\ (\bibinfo  {publisher}
  {Wiley},\ \bibinfo {address} {Hoboken, NJ},\ \bibinfo {year}
  {2012})\BibitemShut {NoStop}%
\bibitem [{\citenamefont {Chang}\ \emph {et~al.}(2020)\citenamefont {Chang},
  \citenamefont {Zhong}, \citenamefont {Satzinger}, \citenamefont {Chou},
  \citenamefont {Bienfait}, \citenamefont {Conner}, \citenamefont {Dumur},
  \citenamefont {Grebel}, \citenamefont {Peairs}, \citenamefont {Povey},\ and\
  \citenamefont {Cleland}}]{Chang2020}%
  \BibitemOpen
  \bibfield  {author} {\bibinfo {author} {\bibfnamefont {H.-S.}\ \bibnamefont
  {Chang}}, \bibinfo {author} {\bibfnamefont {Y.~P.}\ \bibnamefont {Zhong}},
  \bibinfo {author} {\bibfnamefont {K.~J.}\ \bibnamefont {Satzinger}}, \bibinfo
  {author} {\bibfnamefont {M.-H.}\ \bibnamefont {Chou}}, \bibinfo {author}
  {\bibfnamefont {A.}~\bibnamefont {Bienfait}}, \bibinfo {author}
  {\bibfnamefont {C.~R.}\ \bibnamefont {Conner}}, \bibinfo {author}
  {\bibfnamefont {{\'E}.}~\bibnamefont {Dumur}}, \bibinfo {author}
  {\bibfnamefont {J.}~\bibnamefont {Grebel}}, \bibinfo {author} {\bibfnamefont
  {G.~A.}\ \bibnamefont {Peairs}}, \bibinfo {author} {\bibfnamefont {R.~G.}\
  \bibnamefont {Povey}}, \ and\ \bibinfo {author} {\bibfnamefont {A.~N.}\
  \bibnamefont {Cleland}},\ }\href@noop {} {\bibfield  {journal} {\bibinfo
  {journal} {In preparation}\ } (\bibinfo {year} {2020})}\BibitemShut {NoStop}%
\bibitem [{\citenamefont {Steffen}\ \emph {et~al.}(2006)\citenamefont
  {Steffen}, \citenamefont {Ansmann}, \citenamefont {Bialczak}, \citenamefont
  {Katz}, \citenamefont {Lucero}, \citenamefont {McDermott}, \citenamefont
  {Neeley}, \citenamefont {Weig}, \citenamefont {Cleland},\ and\ \citenamefont
  {Martinis}}]{Steffen2006}%
  \BibitemOpen
  \bibfield  {author} {\bibinfo {author} {\bibfnamefont {M.}~\bibnamefont
  {Steffen}}, \bibinfo {author} {\bibfnamefont {M.}~\bibnamefont {Ansmann}},
  \bibinfo {author} {\bibfnamefont {R.~C.}\ \bibnamefont {Bialczak}}, \bibinfo
  {author} {\bibfnamefont {N.}~\bibnamefont {Katz}}, \bibinfo {author}
  {\bibfnamefont {E.}~\bibnamefont {Lucero}}, \bibinfo {author} {\bibfnamefont
  {R.}~\bibnamefont {McDermott}}, \bibinfo {author} {\bibfnamefont
  {M.}~\bibnamefont {Neeley}}, \bibinfo {author} {\bibfnamefont {E.~M.}\
  \bibnamefont {Weig}}, \bibinfo {author} {\bibfnamefont {A.~N.}\ \bibnamefont
  {Cleland}}, \ and\ \bibinfo {author} {\bibfnamefont {J.~M.}\ \bibnamefont
  {Martinis}},\ }\href {\doibase 10.1126/science.1130886} {\bibfield  {journal}
  {\bibinfo  {journal} {Science}\ }\textbf {\bibinfo {volume} {313}},\ \bibinfo
  {pages} {1423} (\bibinfo {year} {2006})}\BibitemShut {NoStop}%
\bibitem [{\citenamefont {Chow}\ \emph {et~al.}(2010)\citenamefont {Chow},
  \citenamefont {DiCarlo}, \citenamefont {Gambetta}, \citenamefont
  {Nunnenkamp}, \citenamefont {Bishop}, \citenamefont {Frunzio}, \citenamefont
  {Devoret}, \citenamefont {Girvin},\ and\ \citenamefont
  {Schoelkopf}}]{Chow2010}%
  \BibitemOpen
  \bibfield  {author} {\bibinfo {author} {\bibfnamefont {J.~M.}\ \bibnamefont
  {Chow}}, \bibinfo {author} {\bibfnamefont {L.}~\bibnamefont {DiCarlo}},
  \bibinfo {author} {\bibfnamefont {J.~M.}\ \bibnamefont {Gambetta}}, \bibinfo
  {author} {\bibfnamefont {A.}~\bibnamefont {Nunnenkamp}}, \bibinfo {author}
  {\bibfnamefont {L.~S.}\ \bibnamefont {Bishop}}, \bibinfo {author}
  {\bibfnamefont {L.}~\bibnamefont {Frunzio}}, \bibinfo {author} {\bibfnamefont
  {M.~H.}\ \bibnamefont {Devoret}}, \bibinfo {author} {\bibfnamefont {S.~M.}\
  \bibnamefont {Girvin}}, \ and\ \bibinfo {author} {\bibfnamefont {R.~J.}\
  \bibnamefont {Schoelkopf}},\ }\href {\doibase 10.1103/PhysRevA.81.062325}
  {\bibfield  {journal} {\bibinfo  {journal} {Physical Review A}\ }\textbf
  {\bibinfo {volume} {81}},\ \bibinfo {pages} {062325} (\bibinfo {year}
  {2010})}\BibitemShut {NoStop}%
\bibitem [{\citenamefont {Neeley}\ \emph {et~al.}(2010)\citenamefont {Neeley},
  \citenamefont {Bialczak}, \citenamefont {Lenander}, \citenamefont {Lucero},
  \citenamefont {Mariantoni}, \citenamefont {O'Connell}, \citenamefont {Sank},
  \citenamefont {Wang}, \citenamefont {Weides}, \citenamefont {Wenner},
  \citenamefont {Yin}, \citenamefont {Yamamoto}, \citenamefont {Cleland},\ and\
  \citenamefont {Martinis}}]{Neeley2010}%
  \BibitemOpen
  \bibfield  {author} {\bibinfo {author} {\bibfnamefont {M.}~\bibnamefont
  {Neeley}}, \bibinfo {author} {\bibfnamefont {R.~C.}\ \bibnamefont
  {Bialczak}}, \bibinfo {author} {\bibfnamefont {M.}~\bibnamefont {Lenander}},
  \bibinfo {author} {\bibfnamefont {E.}~\bibnamefont {Lucero}}, \bibinfo
  {author} {\bibfnamefont {M.}~\bibnamefont {Mariantoni}}, \bibinfo {author}
  {\bibfnamefont {A.~D.}\ \bibnamefont {O'Connell}}, \bibinfo {author}
  {\bibfnamefont {D.}~\bibnamefont {Sank}}, \bibinfo {author} {\bibfnamefont
  {H.}~\bibnamefont {Wang}}, \bibinfo {author} {\bibfnamefont {M.}~\bibnamefont
  {Weides}}, \bibinfo {author} {\bibfnamefont {J.}~\bibnamefont {Wenner}},
  \bibinfo {author} {\bibfnamefont {Y.}~\bibnamefont {Yin}}, \bibinfo {author}
  {\bibfnamefont {T.}~\bibnamefont {Yamamoto}}, \bibinfo {author}
  {\bibfnamefont {A.~N.}\ \bibnamefont {Cleland}}, \ and\ \bibinfo {author}
  {\bibfnamefont {J.~M.}\ \bibnamefont {Martinis}},\ }\href {\doibase
  10.1038/nature09418} {\bibfield  {journal} {\bibinfo  {journal} {Nature}\
  }\textbf {\bibinfo {volume} {467}},\ \bibinfo {pages} {570} (\bibinfo {year}
  {2010})}\BibitemShut {NoStop}%
\bibitem [{\citenamefont {Neeley}\ \emph {et~al.}(2008)\citenamefont {Neeley},
  \citenamefont {Ansmann}, \citenamefont {Bialczak}, \citenamefont {Hofheinz},
  \citenamefont {Katz}, \citenamefont {Lucero}, \citenamefont {O'Connell},
  \citenamefont {Wang}, \citenamefont {Cleland},\ and\ \citenamefont
  {Martinis}}]{Neeley2008}%
  \BibitemOpen
  \bibfield  {author} {\bibinfo {author} {\bibfnamefont {M.}~\bibnamefont
  {Neeley}}, \bibinfo {author} {\bibfnamefont {M.}~\bibnamefont {Ansmann}},
  \bibinfo {author} {\bibfnamefont {R.~C.}\ \bibnamefont {Bialczak}}, \bibinfo
  {author} {\bibfnamefont {M.}~\bibnamefont {Hofheinz}}, \bibinfo {author}
  {\bibfnamefont {N.}~\bibnamefont {Katz}}, \bibinfo {author} {\bibfnamefont
  {E.}~\bibnamefont {Lucero}}, \bibinfo {author} {\bibfnamefont
  {A.}~\bibnamefont {O'Connell}}, \bibinfo {author} {\bibfnamefont
  {H.}~\bibnamefont {Wang}}, \bibinfo {author} {\bibfnamefont {A.~N.}\
  \bibnamefont {Cleland}}, \ and\ \bibinfo {author} {\bibfnamefont {J.~M.}\
  \bibnamefont {Martinis}},\ }\href {\doibase 10.1038/nphys972} {\bibfield
  {journal} {\bibinfo  {journal} {Nature Physics}\ }\textbf {\bibinfo {volume}
  {4}},\ \bibinfo {pages} {523} (\bibinfo {year} {2008})}\BibitemShut {NoStop}%
\bibitem [{\citenamefont {Vitanov}\ \emph {et~al.}(2017)\citenamefont
  {Vitanov}, \citenamefont {Rangelov}, \citenamefont {Shore},\ and\
  \citenamefont {Bergmann}}]{Vitanov2017}%
  \BibitemOpen
  \bibfield  {author} {\bibinfo {author} {\bibfnamefont {N.~V.}\ \bibnamefont
  {Vitanov}}, \bibinfo {author} {\bibfnamefont {A.~A.}\ \bibnamefont
  {Rangelov}}, \bibinfo {author} {\bibfnamefont {B.~W.}\ \bibnamefont {Shore}},
  \ and\ \bibinfo {author} {\bibfnamefont {K.}~\bibnamefont {Bergmann}},\
  }\href {\doibase 10.1103/RevModPhys.89.015006} {\bibfield  {journal}
  {\bibinfo  {journal} {Reviews of Modern Physics}\ }\textbf {\bibinfo {volume}
  {89}},\ \bibinfo {pages} {015006} (\bibinfo {year} {2017})}\BibitemShut
  {NoStop}%
\bibitem [{\citenamefont {Xu}\ \emph {et~al.}(2016)\citenamefont {Xu},
  \citenamefont {Song}, \citenamefont {Liu}, \citenamefont {Xue}, \citenamefont
  {Su}, \citenamefont {Deng}, \citenamefont {Tian}, \citenamefont {Zheng},
  \citenamefont {Han}, \citenamefont {Zhong}, \citenamefont {Wang},
  \citenamefont {Liu},\ and\ \citenamefont {Zhao}}]{Xu2016}%
  \BibitemOpen
  \bibfield  {author} {\bibinfo {author} {\bibfnamefont {H.~K.}\ \bibnamefont
  {Xu}}, \bibinfo {author} {\bibfnamefont {C.}~\bibnamefont {Song}}, \bibinfo
  {author} {\bibfnamefont {W.~Y.}\ \bibnamefont {Liu}}, \bibinfo {author}
  {\bibfnamefont {G.~M.}\ \bibnamefont {Xue}}, \bibinfo {author} {\bibfnamefont
  {F.~F.}\ \bibnamefont {Su}}, \bibinfo {author} {\bibfnamefont
  {H.}~\bibnamefont {Deng}}, \bibinfo {author} {\bibfnamefont {Y.}~\bibnamefont
  {Tian}}, \bibinfo {author} {\bibfnamefont {D.~N.}\ \bibnamefont {Zheng}},
  \bibinfo {author} {\bibfnamefont {S.}~\bibnamefont {Han}}, \bibinfo {author}
  {\bibfnamefont {Y.~P.}\ \bibnamefont {Zhong}}, \bibinfo {author}
  {\bibfnamefont {H.}~\bibnamefont {Wang}}, \bibinfo {author} {\bibfnamefont
  {Y.-x.}\ \bibnamefont {Liu}}, \ and\ \bibinfo {author} {\bibfnamefont
  {S.~P.}\ \bibnamefont {Zhao}},\ }\href {\doibase 10.1038/ncomms11018}
  {\bibfield  {journal} {\bibinfo  {journal} {Nature Communications}\ }\textbf
  {\bibinfo {volume} {7}},\ \bibinfo {pages} {11018} (\bibinfo {year}
  {2016})}\BibitemShut {NoStop}%
\bibitem [{\citenamefont {Bergmann}\ \emph {et~al.}(1998)\citenamefont
  {Bergmann}, \citenamefont {Theuer},\ and\ \citenamefont
  {Shore}}]{Bergmann1998}%
  \BibitemOpen
  \bibfield  {author} {\bibinfo {author} {\bibfnamefont {K.}~\bibnamefont
  {Bergmann}}, \bibinfo {author} {\bibfnamefont {H.}~\bibnamefont {Theuer}}, \
  and\ \bibinfo {author} {\bibfnamefont {B.~W.}\ \bibnamefont {Shore}},\ }\href
  {\doibase 10.1103/RevModPhys.70.1003} {\bibfield  {journal} {\bibinfo
  {journal} {Reviews of Modern Physics}\ }\textbf {\bibinfo {volume} {70}},\
  \bibinfo {pages} {1003} (\bibinfo {year} {1998})}\BibitemShut {NoStop}%
\bibitem [{\citenamefont {Shore}(2011)}]{Shore2011}%
  \BibitemOpen
  \bibfield  {author} {\bibinfo {author} {\bibfnamefont {B.~W.}\ \bibnamefont
  {Shore}},\ }\href@noop {} {\emph {\bibinfo {title} {Manipulating quantum
  structures using laser pulses}}}\ (\bibinfo  {publisher} {Cambridge
  University Press},\ \bibinfo {address} {Cambridge, UK ; New York},\ \bibinfo
  {year} {2011})\BibitemShut {NoStop}%
\bibitem [{\citenamefont {Scully}\ and\ \citenamefont
  {Zubairy}(1997)}]{Scully1997}%
  \BibitemOpen
  \bibfield  {author} {\bibinfo {author} {\bibfnamefont {M.~O.}\ \bibnamefont
  {Scully}}\ and\ \bibinfo {author} {\bibfnamefont {M.~S.}\ \bibnamefont
  {Zubairy}},\ }\href@noop {} {\emph {\bibinfo {title} {Quantum optics}}}\
  (\bibinfo  {publisher} {Cambridge University Press},\ \bibinfo {address}
  {Cambridge ; New York},\ \bibinfo {year} {1997})\BibitemShut {NoStop}%
\bibitem [{\citenamefont {Vasilev}\ \emph {et~al.}(2009)\citenamefont
  {Vasilev}, \citenamefont {Kuhn},\ and\ \citenamefont
  {Vitanov}}]{Vasilev2009}%
  \BibitemOpen
  \bibfield  {author} {\bibinfo {author} {\bibfnamefont {G.~S.}\ \bibnamefont
  {Vasilev}}, \bibinfo {author} {\bibfnamefont {A.}~\bibnamefont {Kuhn}}, \
  and\ \bibinfo {author} {\bibfnamefont {N.~V.}\ \bibnamefont {Vitanov}},\
  }\href {\doibase 10.1103/PhysRevA.80.013417} {\bibfield  {journal} {\bibinfo
  {journal} {Physical Review A}\ }\textbf {\bibinfo {volume} {80}},\ \bibinfo
  {pages} {013417} (\bibinfo {year} {2009})}\BibitemShut {NoStop}%
\bibitem [{\citenamefont {Guerin}\ \emph {et~al.}(2002)\citenamefont {Guerin},
  \citenamefont {Thomas},\ and\ \citenamefont {Jauslin}}]{Guerin2002}%
  \BibitemOpen
  \bibfield  {author} {\bibinfo {author} {\bibfnamefont {S.}~\bibnamefont
  {Guerin}}, \bibinfo {author} {\bibfnamefont {S.}~\bibnamefont {Thomas}}, \
  and\ \bibinfo {author} {\bibfnamefont {H.~R.}\ \bibnamefont {Jauslin}},\
  }\href {\doibase 10.1103/PhysRevA.65.023409} {\bibfield  {journal} {\bibinfo
  {journal} {Physical Review A}\ }\textbf {\bibinfo {volume} {65}},\ \bibinfo
  {pages} {023409} (\bibinfo {year} {2002})}\BibitemShut {NoStop}%
\bibitem [{\citenamefont {Pellizzari}(1997)}]{Pellizzari1997}%
  \BibitemOpen
  \bibfield  {author} {\bibinfo {author} {\bibfnamefont {T.}~\bibnamefont
  {Pellizzari}},\ }\href {\doibase 10.1103/PhysRevLett.79.5242} {\bibfield
  {journal} {\bibinfo  {journal} {Physical Review Letters}\ }\textbf {\bibinfo
  {volume} {79}},\ \bibinfo {pages} {5242} (\bibinfo {year}
  {1997})}\BibitemShut {NoStop}%
\bibitem [{\citenamefont {Vogell}\ \emph {et~al.}(2017)\citenamefont {Vogell},
  \citenamefont {Vermersch}, \citenamefont {Northup}, \citenamefont {Lanyon},\
  and\ \citenamefont {Muschik}}]{Vogell2017}%
  \BibitemOpen
  \bibfield  {author} {\bibinfo {author} {\bibfnamefont {B.}~\bibnamefont
  {Vogell}}, \bibinfo {author} {\bibfnamefont {B.}~\bibnamefont {Vermersch}},
  \bibinfo {author} {\bibfnamefont {T.~E.}\ \bibnamefont {Northup}}, \bibinfo
  {author} {\bibfnamefont {B.~P.}\ \bibnamefont {Lanyon}}, \ and\ \bibinfo
  {author} {\bibfnamefont {C.~A.}\ \bibnamefont {Muschik}},\ }\href {\doibase
  10.1088/2058-9565/aa868b} {\bibfield  {journal} {\bibinfo  {journal} {Quantum
  Science and Technology}\ }\textbf {\bibinfo {volume} {2}},\ \bibinfo {pages}
  {045003} (\bibinfo {year} {2017})}\BibitemShut {NoStop}%
\bibitem [{\citenamefont {Lindblad}(1976)}]{Lindblad1976}%
  \BibitemOpen
  \bibfield  {author} {\bibinfo {author} {\bibfnamefont {G.}~\bibnamefont
  {Lindblad}},\ }\href {\doibase 10.1007/BF01608499} {\bibfield  {journal}
  {\bibinfo  {journal} {Communications in Mathematical Physics}\ }\textbf
  {\bibinfo {volume} {48}},\ \bibinfo {pages} {119} (\bibinfo {year}
  {1976})}\BibitemShut {NoStop}%
\bibitem [{\citenamefont {Walls}\ and\ \citenamefont
  {Milburn}(2008)}]{Walls2008}%
  \BibitemOpen
  \bibfield  {author} {\bibinfo {author} {\bibfnamefont {D.~F.}\ \bibnamefont
  {Walls}}\ and\ \bibinfo {author} {\bibfnamefont {G.~J.}\ \bibnamefont
  {Milburn}},\ }\href@noop {} {\emph {\bibinfo {title} {Quantum optics}}},\
  \bibinfo {edition} {2nd}\ ed.\ (\bibinfo  {publisher} {Springer},\ \bibinfo
  {address} {Berlin},\ \bibinfo {year} {2008})\BibitemShut {NoStop}%
\bibitem [{\citenamefont {Johansson}\ \emph {et~al.}(2012)\citenamefont
  {Johansson}, \citenamefont {Nation},\ and\ \citenamefont
  {Nori}}]{Johansson2012}%
  \BibitemOpen
  \bibfield  {author} {\bibinfo {author} {\bibfnamefont {J.}~\bibnamefont
  {Johansson}}, \bibinfo {author} {\bibfnamefont {P.}~\bibnamefont {Nation}}, \
  and\ \bibinfo {author} {\bibfnamefont {F.}~\bibnamefont {Nori}},\ }\href
  {\doibase 10.1016/j.cpc.2012.02.021} {\bibfield  {journal} {\bibinfo
  {journal} {Computer Physics Communications}\ }\textbf {\bibinfo {volume}
  {183}},\ \bibinfo {pages} {1760} (\bibinfo {year} {2012})}\BibitemShut
  {NoStop}%
\bibitem [{\citenamefont {Wang}\ \emph {et~al.}(2017)\citenamefont {Wang},
  \citenamefont {Zhang}, \citenamefont {Yan},\ and\ \citenamefont
  {Chesi}}]{Wang2017}%
  \BibitemOpen
  \bibfield  {author} {\bibinfo {author} {\bibfnamefont {Y.-D.}\ \bibnamefont
  {Wang}}, \bibinfo {author} {\bibfnamefont {R.}~\bibnamefont {Zhang}},
  \bibinfo {author} {\bibfnamefont {X.-B.}\ \bibnamefont {Yan}}, \ and\
  \bibinfo {author} {\bibfnamefont {S.}~\bibnamefont {Chesi}},\ }\href
  {\doibase 10.1088/1367-2630/aa7f5d} {\bibfield  {journal} {\bibinfo
  {journal} {New Journal of Physics}\ }\textbf {\bibinfo {volume} {19}},\
  \bibinfo {pages} {093016} (\bibinfo {year} {2017})}\BibitemShut {NoStop}%
\bibitem [{\citenamefont {Wootters}(1998)}]{Wootters1998}%
  \BibitemOpen
  \bibfield  {author} {\bibinfo {author} {\bibfnamefont {W.~K.}\ \bibnamefont
  {Wootters}},\ }\href {\doibase 10.1103/PhysRevLett.80.2245} {\bibfield
  {journal} {\bibinfo  {journal} {Physical Review Letters}\ }\textbf {\bibinfo
  {volume} {80}},\ \bibinfo {pages} {2245} (\bibinfo {year}
  {1998})}\BibitemShut {NoStop}%
\bibitem [{\citenamefont {Plenio}\ and\ \citenamefont
  {Virmani}(2006)}]{Plenio2006}%
  \BibitemOpen
  \bibfield  {author} {\bibinfo {author} {\bibfnamefont {M.~B.}\ \bibnamefont
  {Plenio}}\ and\ \bibinfo {author} {\bibfnamefont {S.}~\bibnamefont
  {Virmani}},\ }\href@noop {} {\bibfield  {journal} {\bibinfo  {journal}
  {arXiv:quant-ph/0504163}\ } (\bibinfo {year} {2006})},\ \bibinfo {note}
  {arXiv: quant-ph/0504163}\BibitemShut {NoStop}%
\end{thebibliography}%

\end{document}

\begin{center}
	\textbf{\Large{Supplementary Materials for Remote entanglement via adiabatic passage using a tunably-dissipative quantum communication system}}
\end{center}

\setcounter{figure}{0}   
\renewcommand{\thefigure}{S\arabic{figure}}
\renewcommand{\thetable}{S\arabic{table}}
\renewcommand{\theequation}{S\arabic{equation}}

\section{Device and experimental setup}
\begin{figure}[H]
	\centering		
	\includegraphics[width=\textwidth]{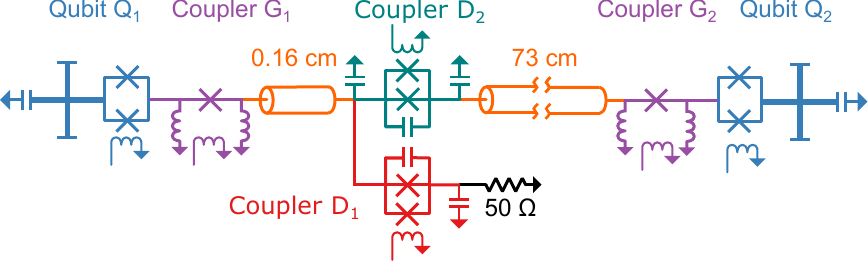}
	\caption{
		\label{fig:s_circuit}
		Circuit diagram for the experimental device.
		The qubits are in blue, their tunable couplers in purple, the two couplers making up the switch in teal and red, the transmission line in orange, and the $50~\Omega$ load in black.
	}
	\centering
\end{figure}

The experiment is carried out inside a dilution refrigerator with a base temperature below 10 mK. A detailed description of the experimental setup, as well as the process flow for the device fabrication, are provided in ref.~\onlinecite{Zhong2019}. A circuit diagram is shown in Fig.~\ref{fig:s_circuit} with detailed device parameters provided in Table~\ref{table:parameters}.

\subsection{Superconducting qubits}
The superconducting qubits used in this experiment are frequency-tunable planar transmons \cite{Koch2007,Barends2013}. Microwave lines capacitively-coupled to each qubit are used to generate qubit rotations about the $X$ and $Y$ axes of the Bloch sphere; $Z$-axis rotations and frequency tuning of each qubit are controlled using dc flux-bias lines inductively-coupled to each qubit's two-Josephson junction SQUID loop. To prevent spurious cross-excitations between the two qubits, the qubits are typically de-tuned from one another by 85 MHz, and each qubit's coupler $G_1$ ($G_2$) is turned off during qubit state preparation and readout. Each qubit's intrinsic qubit lifetime, coherence time, and idle frequency are provided in Table~\ref{table:parameters}. Each qubit is read out simultaneously with the other qubit, using a dispersive single-shot readout \cite{Jeffrey2014,Kelly2015} via a capacitively-coupled quarter-wave coplanar waveguide resonator. We used a traveling-wave parametric amplifier \cite{Macklin2015} (MIT Lincoln Laboratories) to ensure nearly quantum-limited amplification of the readout signals. The $|g \rangle$ and $|e\rangle$ state readout fidelities for each qubit are shown in Table~\ref{table:parameters}.

\subsection{Flux-tunable couplers}
The tunable coupling between each qubit and the communication channel is controlled via a galvanically-connected variable coupling $\pi$-bridge \cite{Chen2014,Zhong2019}, labeled as $G_1$ and $G_2$ in Fig.~\ref{fig:s_circuit}. A dc flux-bias line affords flux control of each coupler by changing its Josephson junction inductance. However, changes in the coupler junction inductance induces a sympathetic frequency shift in the qubit connected to that coupler, as the inductance of the qubit is modified as well. Similarly, changes in the coupler junction inductance also shifts the transmission line frequency.  We calibrate these frequency shifts for all qubit frequencies and coupling strengths used in the adiabatic transfer process. This ensures that the qubits and the channel mode remain in frequency resonance during the transfer, and that the couplings $g_1$ and $g_2$ vary precisely according to the desired sine and cosine forms described in the main text. We further note that our choice of total transfer time $t_f=132$~ns corresponds to control pulses with $<2$~MHz bandwidth, far below the maximum bandwidth of our control electronics (250 MHz).

\subsection{Communication channel}
The communication channel connecting the two qubits comprises a 0.73 m-long, on-chip coplanar waveguide. To suppress unwanted slotline modes, the transmission line is spanned by air-bridge crossovers every 2 mm, connecting the ground planes on either side of the transmission line \cite{Zhong2019}. Each resonant standing mode $n$ in the approximately short-circuited line can be modeled as a series $RLC$ resonant circuit with the equivalent lumped-element parameters \cite{Pozar}.
\begin{linenomath}
	\begin{align}
	R_{n} &= Z_0 \alpha \ell, \\
	L_{n} &= \frac{1}{2} \mathscr{L} \ell, \\
	C_{n} &= \frac{1}{\omega_n^2 L_n},
	\end{align}
\end{linenomath}
where $Z_0 = 50~\Omega$ is the characteristic impedance of the line, determined by geometry and substrate, $\alpha = 0.010$~dB/m is the (lossy) real part of the propagation parameter, determined from the intrinsic resonant mode lifetime $T_{1r,\mathrm{int}}$, $\mathscr{L} = 402$~nH/m is the inductance per unit length, $\ell = 0.73$ m is the total length, and $\omega_n = n \omega_{FSR} = n \times 2 \pi~84$ MHz is the resonant frequency of the $n$th standing mode.
\clearpage
\subsection{Tunable switch}
\begin{figure}[h]
	\centering		
	\includegraphics[width=8.6cm]{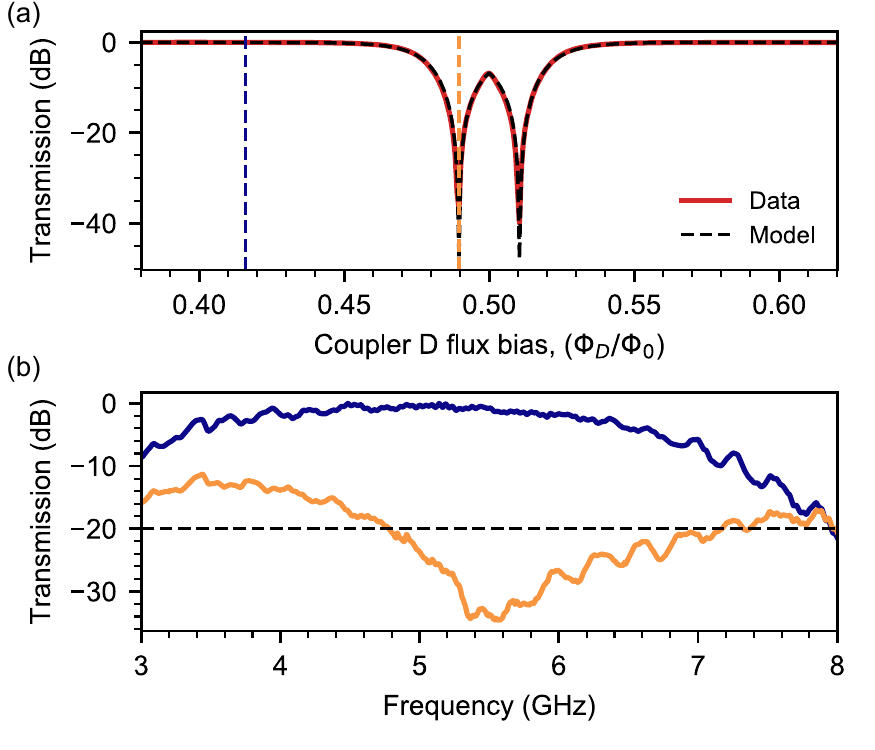}
	\caption{
		\label{fig:s_dcoupler}
		Characterization of a DC SQUID tunable switch, designed identically to those used in the experiments in the main text.
		(a) Transmission through the SQUID tunable switch as a function of flux bias. The on and off flux settings are marked by blue and orange dashed vertical lines. Dashed lines are results from a circuit model.
		(b) Transmission measured as a function of frequency near 5.5 GHz for the on (blue) and off (orange) SQUID settings. These demonstrate an on/off ratio greater than 35 dB, isolation bandwidth (below -20 dB) of about 2.9 GHz, and transmission bandwidth (above -1 dB) larger than 1 GHz. The dashed line is the -20 dB transmission threshold used to define the isolation bandwidth.
	}
	\centering
\end{figure}

The tunable switch placed near qubit $Q_1$ and its tunable coupler $G_1$ consists of two couplers, $D_1$ and $D_2$, each comprising a DC SQUID in line with each branch of the network. These are used to control the flow of microwave signals through the coupler network. SQUID $D_1$ connects to an off-chip $50~\Omega$ load via a wire bond connection, yielding a variable dissipative cold load to the system, while SQUID $D_2$ connects to the transmission line leading to the tunable coupler $G_2$ and qubit $Q_2$. When the SQUID plasma frequency is close to resonance with an incoming signal, the SQUID presents a high-impedance load that almost completely reflects the signal, while when the SQUID frequency is tuned well away from the signal frequency, nearly unit transmission is achieved. Independent experiments on identically-designed SQUID circuits were used to measure signal transmission though the SQUID as a function of tuning flux, and demonstrate greater than 1 GHz bandwidth with on/off ratios in excess of 35 dB \cite{Chang2020}. The transmission dependence on bias flux as well as its frequency dependence for a typical SQUID are shown in Fig.~\ref{fig:s_dcoupler}.

\section{Quantum state and process tomography}
\subsection{Readout correction}
The qubit readout fidelities are displayed in Table~\ref{table:parameters}. These are measured by preparing each qubit in $|g \rangle$ or $|e \rangle$ and performing measurements in the two-qubit basis, $|gg \rangle, |ge \rangle, |eg \rangle$ and  $|ee \rangle$. These yield an assignment probability matrix, which is used for readout error correction through linear inversion \cite{Steffen2006,Chow2010}. A typical assignment probability matrix is shown in Eq.~\ref{eq.assign}. In the main text, we display the qubit excited state populations, and the quantum process and state tomography fidelities, which are all corrected for measurement errors. As shown in Table~\ref{table:qpt}, there is a modest difference between fidelities obtained with or without these readout corrections.
\begin{linenomath}
	\begin{align}\label{eq.assign}
	M=\begin{bmatrix}
	0.926 & 0.107 & 0.114 & 0.013 \\
	0.040 & 0.865 & 0.005 & 0.120 \\
	0.033 & 0.005 & 0.853 & 0.107 \\
	0.001 & 0.023 & 0.028 & 0.759
	\end{bmatrix}
	\end{align}
\end{linenomath}
\subsection{Quantum state tomography}
We carry out quantum state tomography by applying the single tomography gates $\left \lbrace I, R_x^{\pi / 2}, R_y^{\pi / 2} \right \rbrace$ and then reading out both qubits simultaneously. The density matrix is reconstructed using linear inversion to correct for measurement error and validated to ensure the resulting density matrix $\rho$ is Hermitian, positive, and semi-definite with unit trace \cite{Steffen2006, Neeley2010}. In the experiment, $Q_2$'s tomography pulse is rotated by a calibrated azimuthal angle $\varphi$ on the Bloch sphere to account for the phase accumulated from the relative detunings of the two qubits during the transfer sequence.

\subsection{Quantum process tomography}
We perform quantum process tomography by preparing four representative single-qubit input states at the sending qubit, $\left \lbrace | g
\rangle, (| g \rangle + | e \rangle) / \sqrt{2}, (| g \rangle + i | e \rangle)/ \sqrt{2}, | e \rangle \right \rbrace$, and subsequently carrying out the state transfer protocol. At the end of the transfer, we measure the resulting density matrix for the receiver qubit via quantum state tomography, and we calculate the process fidelity through linear inversion. The process matrix is validated to ensure that it is positive, Hermitian, and semi-definite with unit trace \cite{Neeley2008}. In Table~\ref{table:qpt}, we show the process fidelities and trace distances obtained using the adiabatic protocol for the six dissipation settings explored in the main text.

\begin{table}[h]
	\begin{center}
		\begin{tabular}{l c c c c c}
			\hline
			$T_{1r}$ (ns) & Fidelity & Fidelity (corrected)   &  Trace distance \\
			& ${\mathcal F}_m$ & ${\mathcal F}_c$ & ${\mathcal D}$ \\
			\hline
			$28.7 \pm 0.2$ & $0.77 \pm 0.01$ & $0.79 \pm 0.01$ & 0.05 \\
			$49.8 \pm 0.3$ & $0.80 \pm 0.01$ & $0.83 \pm 0.01$ & 0.06 \\
			$101.1 \pm 0.7$ & $0.86 \pm 0.01$ & $0.87 \pm 0.01$ & 0.03 \\
			$336 \pm 3$ & $0.91 \pm 0.01$ & $0.92 \pm 0.01$ & 0.03 \\
			$503 \pm 5$ & $0.92 \pm 0.01$ & $0.93 \pm 0.01$ & 0.02 \\
			$3410 \pm 40$ & $0.93 \pm 0.01$ & $0.96 \pm 0.01$ & 0.02 \\
			\hline
		\end{tabular}
	\end{center}
	\caption{\label{table:qpt}
		Quantum process tomography for adiabatic state transfer at each dissipation level in the channel described in the main text. The measured fidelity is calculated from ${\mathcal F}_m = \mathrm{Tr}(\chi_m \cdot \chi_{\mathrm{ideal}})$, where $\chi_m$ is the process matrix without measurement correction, and the measurement-corrected fidelity ${\mathcal F}_c = \mathrm{Tr}(\chi_c \cdot \chi_{\mathrm{ideal}})$, where $\chi_c$ is corrected for readout error. The trace distance is calculated from ${\mathcal D} = \sqrt{\mathrm{Tr}\Big ([\chi_c -\chi_{\mathrm{sim}}]^2\Big )}$.}
\end{table}

\section{Theory of adiabatic state transfer}
\subsection{State transfer via the dark state}
We present here the theory for the adiabatic protocol implemented in the experiments described in the main text. We assume the three quantum systems (qubit $Q_1$, the transmission line standing mode, and qubit $Q_2$), are all frequency-resonant, and we restrict the discussion to the single-excitation subspace of this system. We can write the relevant terms in the system Hamiltonian in the rotating frame of the coupled system as
\begin{linenomath}
	\begin{align}\label{cpl}
	H/\hbar = g_{1}(t)(|e0g \rangle \langle g1g| + |g1g \rangle \langle e0g|) + g_{2}(t) (|g0e \rangle \langle g1g| + |g1g \rangle \langle g0e|),
	\end{align}
\end{linenomath}
where $g_1(t)$ is the time-dependent coupling between qubit $Q_1$ and the transmission line standing mode, and $g_2(t)$ that for qubit $Q_2$.

Diagonalizing the Hamiltonian reveals three instantaneous eigenstates of the coupled system:
\begin{linenomath}
	\begin{align}\label{eigenstate}
	|B_{\pm} (t)\rangle &= \frac{1}{\sqrt{2}} \left (\sin\theta(t)|e0g \rangle + \cos\theta(t)|g0e \rangle \pm |g1g\rangle \right ), \\
	|D(t)\rangle &= \cos\theta(t) |e0g \rangle - \sin \theta(t) |g0e\rangle,
	\end{align}
\end{linenomath}
where the instantaneous mixing angle $\theta(t)$ is given by
\begin{linenomath}
	\begin{equation}
	\tan \theta(t) = g_1(t)/g_2(t).
	\end{equation}
\end{linenomath}
The ``dark'' eigenstate $|D(t) \rangle$ has no occupancy in the transmission line mode and is at zero energy. The two eigenstates  $|B_{\pm}(t) \rangle$ are the so-called ``bright'' states, as they include photon occupancy of the transmission line mode. These states have the eigenenergies $E_{\pm} = \pm \hbar \bar{g}$ respectively, where $\bar{g} = \sqrt{g_1(t)^2 + g_2(t)^2}$.

\begin{figure}[h]
	\centering		
	\includegraphics[width=\textwidth]{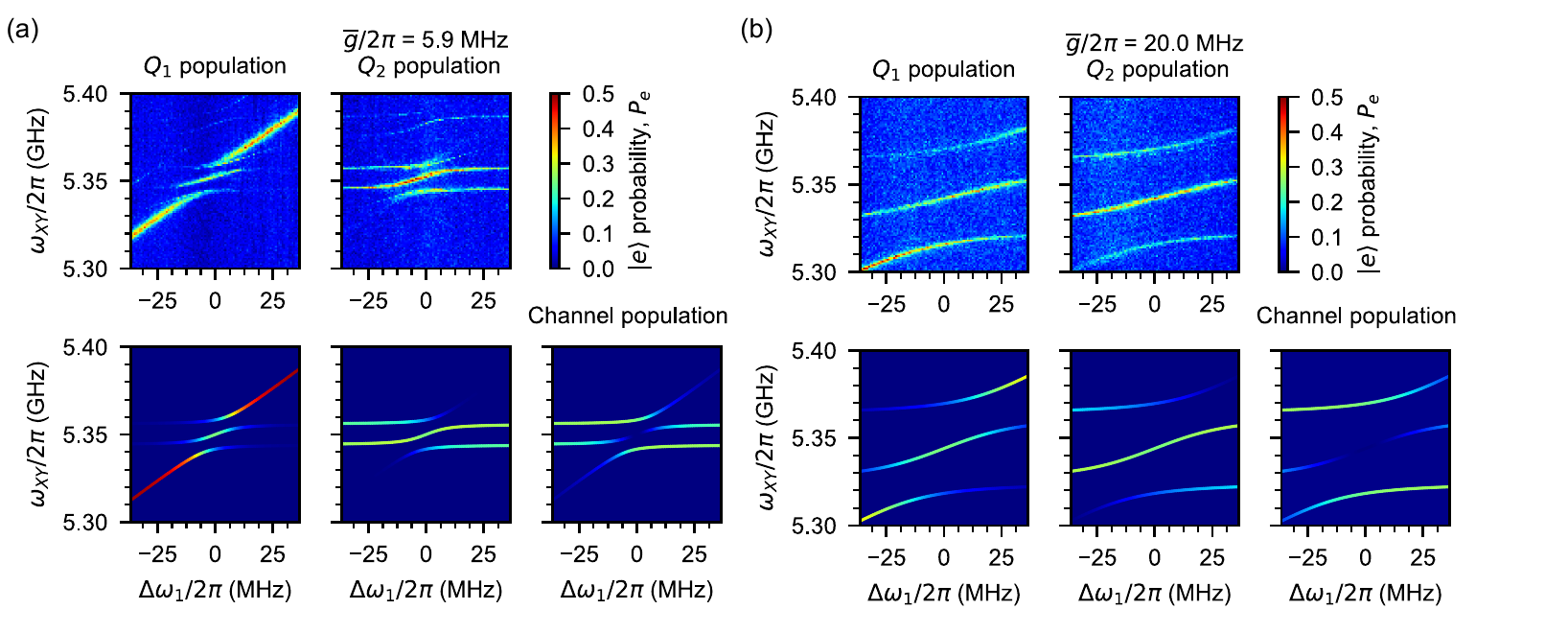}
	\caption{
		\label{fig:s_cplspec}
		Two-qubit coupled spectroscopy near the resonant channel mode $\omega_{\mathrm{r}}/2\pi = 5.351$ GHz at two coupler settings,
		(a) $g_1/2\pi = g_2/2\pi = 5.9\pm0.1$ MHz and
		(b) $g_1/2\pi = g_2/2\pi = 20.0\pm0.1$ MHz. Upper panels are experimental measurements, lower panels are numerical simulations. $Q_2$ is set to be resonant with the channel mode and $Q_1$ is biased to frequency $\omega_{\mathrm{r}} + \Delta \omega_1$, where $\Delta \omega_1$ is varied along the horizontal axis. Qubit spectroscopy is carried out by driving $Q_1$ with a weak 5 $\mu$s-long pulse at each frequency $\omega_{XY}/2 \pi$, then simultaneously measuring each qubit's excited state population $P_{e,Q_1}$ and $P_{e,Q_2}$ using dispersive readout \cite{Jeffrey2014,Kelly2015}. The two bright states are frequency-offset from the zero-energy dark eigenstate by the coupling $\pm \bar{g}/2\pi = \pm \sqrt{g_1^2+g_2^2}/2\pi$.
	}
	\centering
\end{figure}

The dressed eigenstates can be revealed using qubit spectroscopy. In Fig.~\ref{fig:s_cplspec}, with $Q_2$ resonant with the channel mode and with fixed couplings $g_1 = g_2$, sweeping $Q_1$'s frequency through the channel mode frequency reveals three eigenstates separated in frequency by $g_{1,2}/2\pi$, as expected. A numerical simulation (Fig.~\ref{fig:s_cplspec}b) correctly identifies the middle eigenstate as the dark state $|D(t) \rangle$, with no occupancy in the channel, with the other two eigenstates above and below $|D(t)\rangle$ identified as the two bright states $|B_{\pm}(t) \rangle$.

The adiabatic protocol uses the dark state $|D(t) \rangle$ to achieve the desired state transfer from $Q_1$ to $Q_2$ without populating the channel mode. This is achieved by using the sine and cosine time dependence for $g_1$ and $g_2$ respectively, as described in the main text, such that the dark state is $|e0g \rangle$ at $t=0$ and $|g0e \rangle$ at $t=t_f$, and varies smoothly between these limits during the transfer.

\subsection{Adiabatic condition}
As the adiabatic protocol relies on remaining in the dark eigenstate throughout the transfer, the protocol needs to be executed slowly, to minimize non-adiabatic errors from coupling to the bright eigenstates. We control for this here by ensuring that integral of the two coupling functions in time satisfies \cite{Vitanov2017, Xu2016, Bergmann1998, Shore2011, Scully1997, Vasilev2009}
\begin{linenomath}
	\begin{align}
	\int_{0}^{t_f} \bar{g}(t) ~\mathrm{d}t = \int_{0}^{t_f} \sqrt{g_1^2 + g_2^2}~\mathrm{d}t \approx 4 \pi,
	\end{align}
\end{linenomath}
which is much greater than the usual minimum threshold of $3 \pi/2$ for efficient state transfer with greater than $85\%$ efficiency \cite{Vitanov2017}.

We note that the simple coupling scheme adopted here keeps the effective coupling $\bar{g} = \sqrt{g_1^2 +g_2^2}$ constant, and correspondingly the energy splittings between the eigenstates are constant during the transfer. This type of coupling scheme is known as a parallel adiabatic passage (PAP) and is commonly adopted in STIRAP-like adiabatic protocol, as non-adiabatic errors are minimized by avoiding anti-level crossing points during the transfer \cite{Vasilev2009, Guerin2002}.

\section{Master equation model}
We model the quantum behavior of the coupled system using the multi-mode Jaynes-Cummings Hamiltonian $H$.  Our simulation model comprises two qubits (lowering operators $\sigma_{1}, \sigma_{2}$) coupled to $2N+1$ harmonic oscillator modes (lowering operators $a_n$). We can write the coupled Hamiltonian in the rotating frame of the resonant channel mode as
\begin{linenomath}
	\begin{align}\label{H}
	H/\hbar = &\Delta\omega_{1} \sigma_{1}^\dag \sigma_{1} + \Delta\omega_{2} \sigma_{2}^\dag \sigma_{2} + \sum_{n=-N}^{N} \Delta_{n} a_n^\dag a_n \\
	+ &\sum_{n = -N}^{N} g_{1}(t) \left (\sigma_{1} a_n^\dag + \sigma_{1}^\dag a_n \right ) + \sum_{n = -N}^{N} g_{2}(t) (-1)^n \left (\sigma_{2} a_n^\dag + \sigma_{2}^\dag a_n \right ),
	\end{align}
\end{linenomath}
where $\Delta\omega_{1,2}$ are the qubit detunings from the central resonant mode $n=0$, $\Delta_n = n \omega_{\mathrm{FSR}}$ is the detuning of the $n$th channel mode from the $n=0$ central mode, and $g_{1}(t)$ and $g_{2}(t)$ are the time-dependent couplings of $Q_1$ and $Q_2$ to the $n$th channel mode, assumed to be independent of $n$. This is justified by the high mode number ($\sim64$) of the resonant channel modes used; neighboring modes thus have similar coupling strength. We further note that even and odd channel modes have different signs for $g_2$ compared to $g_1$, owing to the parity of their wavefunctions $\psi_{n}(x)$\cite{Pellizzari1997,Vogell2017}.

To simulate the time-domain evolution of our coupled quantum system, we numerically integrate the Lindblad master equation\cite{Lindblad1976,Walls2008} with the Hamiltonian using the python package QuTiP \cite{Johansson2012}. We account for qubit relaxation and decoherence by including the Lindblad collapse operators $\sigma_{-}/\sqrt{T_{1,\mathrm{int}}}$ and $\sigma_{z}/\sqrt{2 T_{\phi}}$, where $1/T_{\phi}=1/T_{2,\mathrm{Ramsey}}-1/2T_{1,\mathrm{int}}$. The energy lifetime of the channel modes $T_{1r}$ is taken to be identical for all oscillator modes and is accounted for by the Lindblad collapse operators $a_n/\sqrt{T_{1r}}$. Qubit parameters are obtained from independent qubit measurements, while $T_{1r}$ is obtained using the method outlined in Fig.~2 of the main text. The numerical simulations include $2N+1 = 5$ modes, each containing two Fock states $|0\rangle$ and $|1\rangle$. The coupling functions $g_{1,2}(t)$ are varied dynamically in time using the coupling described in Fig.~3 of the main text. We use this model to simulate the time evolution of $Q_1$ and $Q_2$ in Fig.~3 of the main text as well as to obtain the expected process and Bell state fidelities, which account for the finite qubit lifetime and coherence (Fig.~3, 4).

\section{Adiabatic protocol in the strong multi-mode coupling regime}
\begin{figure}[h]
	\centering
	\includegraphics[width=8.6cm]{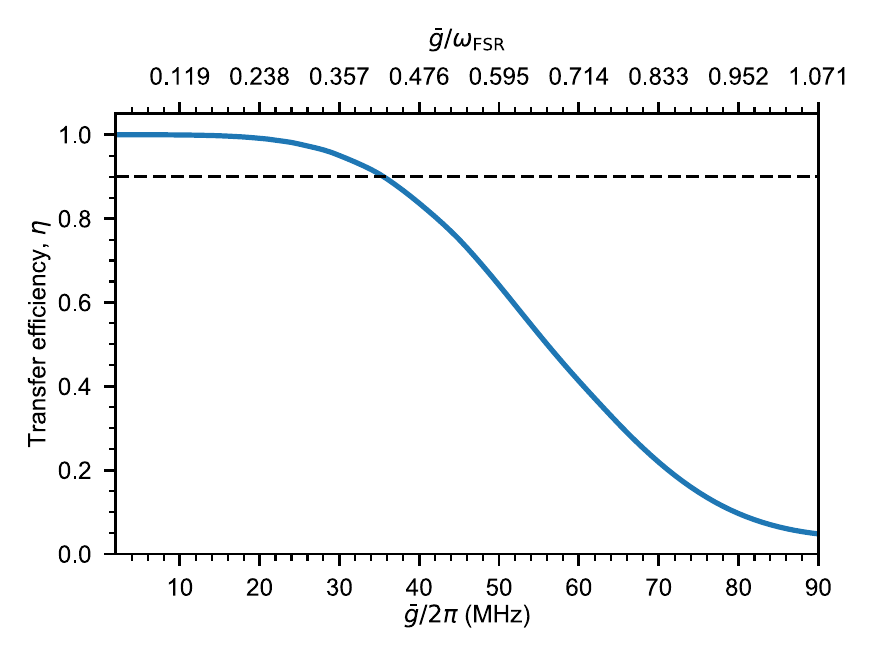}
	\caption{
		\label{fig:s_strongcpl}
		Calculated maximum transfer efficiency $\eta$ as a function of the coupling strength $\bar{g}$. In the numerical simulation, the free spectral range of the channel is kept fixed at $\omega_{FSR}/2\pi=84$~MHz, while the effective coupling strength $\bar{g}$ is varied. For coupling strengths $\bar{g}/2\pi \gtrsim 36$~MHz, interference effects from interactions with neighboring resonant modes become significant, reducing the transfer efficiency attainable with the adiabatic protocol. Dashed line marks where $\eta=90\%$.
	}
	\centering
\end{figure}

Using the master equation model (see above), we explore the performance of our adiabatic protocol as it approaches the strong multi-mode coupling regime, where the coupling between the qubit and the channel mode is of order the free spectral range ($\bar{g} \sim \omega_{FSR}$). We quantify the performance of the protocol by calculating the maximum transfer efficiency $\eta$ attainable at each effective coupling $\bar{g}$. The results of the simulations are shown in Fig~\ref{fig:s_strongcpl}. This simulation includes $2N+1=15$ channel modes, each containing two Fock states $|0\rangle$ and $|1\rangle$.  We did not perform numerical simulations for $\bar{g}/\omega_{FSR} > 1$, as this requires including more than 17 channel modes in the coupled Hamiltonian in Eq.~(\ref{H}) for accurate simulations, consuming significant computational resources for the resultingly large Hilbert space.

\section{Spurious coupling of $Q_1$ to the external load}
\begin{figure}[t]
	\centering	
	\includegraphics[width=11.7cm]{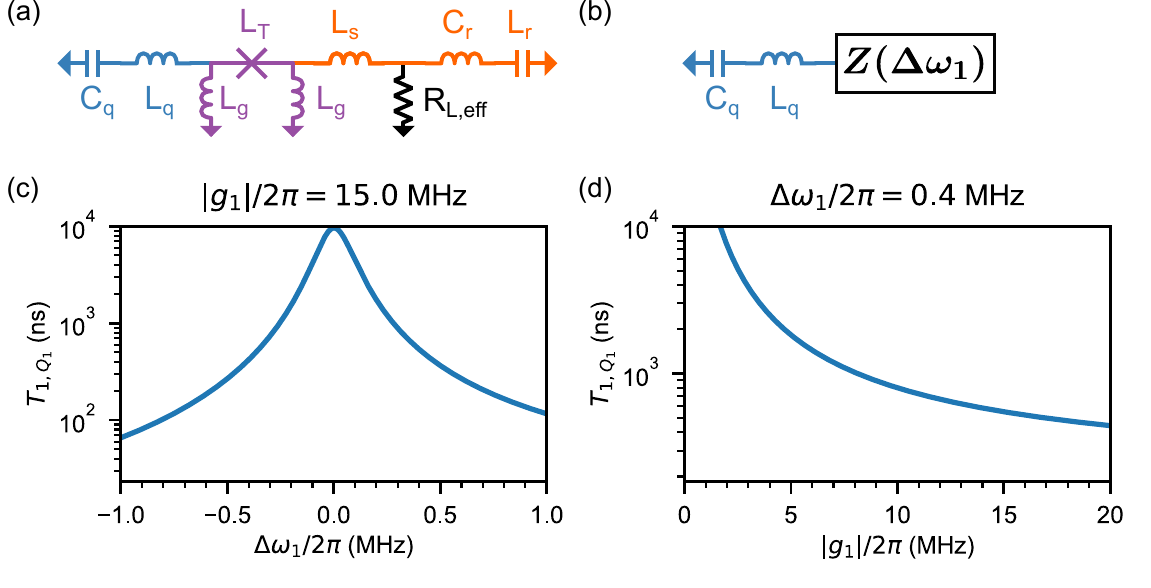}
	\caption{
		\label{fig:s_loadedQ}
		(a) Electrical circuit for calculating the parasitic loading of $Q_1$ from the external $50~\Omega$ load.
		The qubit is represented by the series $C_q-L_q$; the coupler by the $\pi$ bridge circuit $L_g-L_T-L_g$; the short length of transmission line to the load $R_{L,\mathrm{eff}}$ by $L_s$; and finally the lumped model for the channel resonant mode is represented by the series $L_r-C_r$. We then transform the right-half of the circuit to an equivalent impedance $Z(\Delta\omega_1)$ as seen by the qubit (b). We use this circuit model to calculate the loaded energy relaxation times of $Q_1$ as a function of both detuning from the channel mode  $\Delta \omega_1$ and coupling $|g_1|$ using circuit parameters listed in Table~\ref{table:parameters}.
		(c) Calculated $Q_1$ relaxation times as a function of detunings from the resonant mode for the largest dissipation case ($T_{1r} = 28.7$ ns) and with coupling $|g_1|/2\pi=15$~MHz.
		(d) Calculated $Q_1$ relaxation times as a function of coupling $|g_1|$ assuming a constant detuning of 0.4 MHz from the resonant mode.
	}
	\centering
\end{figure}

The primary source of infidelity for the adiabatic protocol is the reduced lifetime of $Q_1$ when the couplers $G_1$ and $D_1$ are both turned on, as this couples both the channel mode and the qubit to the external $50~\Omega$ load. In the ideal case, this coupler only changes the loss in the channel; however, due to the close proximity of $Q_1$ to this coupler in the circuit, the qubit is also be coupled to the $50~\Omega$ load. This can be understood by the simplified circuit model shown in Fig.~\ref{fig:s_loadedQ}a: When $Q_1$ is exactly resonant with the channel mode, the series resonance presented by the channel (represented by the series $L_r-C_r$ in the diagram) shorts the parallel load resistance $R_{L,\textrm{eff}}$, so there is little to no effect on the qubit. Conversely, a slight detuning of the qubit from this resonant frequency increases the $L_r-C_r$ impedance, so the external load is no longer exactly shorted and can load the qubit. This substantially reduces $Q_1$'s $T_1$ lifetime when the coupler to the load is turned on. We model this effect by first calculating the effective external load $R_{L,\textrm{eff}}$ at each dissipation settings in the channel mode
\begin{linenomath}
	\begin{align}\label{R}
	\frac{1}{T_{1r,\textrm{ext}}} &= \frac{1}{T_{1r}} - \frac{1}{T_{1r,\mathrm{int}}}\\
	R_{L,\textrm{eff}} &= \frac{L_r}{T_{1r,\textrm{ext}}}
	\end{align}
\end{linenomath}
Next, we calculate the equivalent impedance $Z(\Delta\omega_1)$ as seen by the qubit as a function of detuning from the channel mode (Fig.~\ref{fig:s_loadedQ}b). The loaded qubit lifetime $T_1$ is then given by:
\begin{linenomath}
	\begin{align}
	T_{1} = L_q/\textrm{Re}[Z(\Delta\omega_1)]
	\end{align}
\end{linenomath}
In Fig.~\ref{fig:s_loadedQ}c,d, we show the calculated energy relaxation time $T_1$ of $Q_1$ due parasitic coupling to the external load at the largest loss case explored here ($T_{1r} = 28.7$ ns) using circuit parameters listed in Table~\ref{table:parameters}. In Fig.~\ref{fig:s_loadedQ}c, we see that for the coupling $|g_1|/2 \pi=15$ MHz, a 0.4 MHz frequency detuning can reduce $Q_1$'s $T_1$ to 500 ns. We further show the coupling strength dependence of this effect assuming a constant detuning in Fig.~\ref{fig:s_loadedQ}d. The relaxation of $Q_1$ for each dissipation setting due to this parasitic coupling has been included in the simulation

\begin{figure}[t]
	\centering		
	\includegraphics[width=8.6cm]{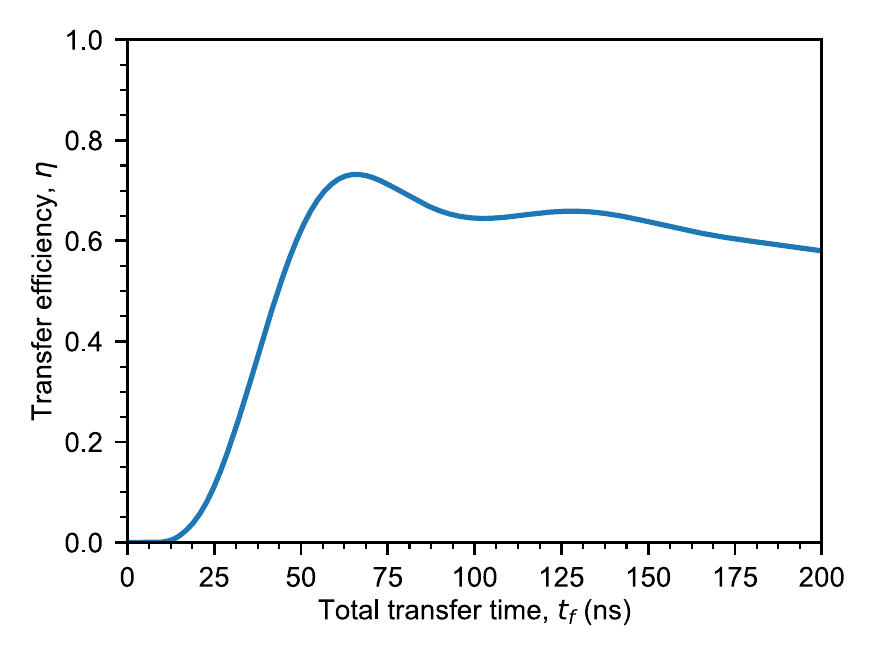}
	\caption{
		\label{fig:s_opt_tf}
		Calculated maximum transfer efficiency as a function of transfer time $t_f$ for the largest loss case explored in the experiment, where $T_{1r} = 28.7$ ns. A maximum transfer efficiency of 0.73 occurs at transfer time of $t_f=66$~ns, 0.06 higher than the efficiency achieved in the experiment with $t_f=132$~ns.
	}
	\centering
\end{figure}

A possible way to overcome this non-ideality and increase the transfer efficiency of the adiabatic protocol further is to decrease the total transfer time $t_f$, reducing the impact of loss from $Q_1$. However, this comes at the cost of populating the channel mode during the transfer, as a result of the reduced adiabaticity. We explore these trade-offs for the largest dissipation case explored here using the master equation model with actual device parameters outlined in Table~\ref{table:parameters}. In Fig.~\ref{fig:s_opt_tf}, we show that a maximum transfer efficiency of $\eta=0.73$ is possible with a $t_f=66$~ns, 0.06 higher than the efficiency achieved in the experiment in the largest loss case, where $T_{1r} = 28.7$ ns, with a total transfer time of $t_f=132$~ns. We also note that in Fig.~\ref{fig:s_opt_tf}, our choice of $t_f=132$~ns in the experiment is a local maximum; this is not coincidental and is expected from theory. The time corresponds to the periodic return of the dark state at discrete times $t_f = (2\pi/\bar{g})\sqrt{n^2-(1/4)^2}$ for non-zero integer $n$ \cite{Wang2017}. For $\bar{g}/2\pi=15$~MHz, our choice of total transfer time $t_f=132$~ns is the $n=2$ case.

\clearpage
\section{Concurrence}
The two-qubit concurrence $\mathcal{C}$ of the Bell singlet state is calculated from the reconstructed density matrix $\rho$ using the standard definition \cite{Wootters1998,Plenio2006}:
\begin{linenomath}
	\begin{align}\label{con}
	\mathcal{C}(\rho) \equiv \mathrm{max} \{0, \lambda_1 -\lambda_2 - \lambda_3 - \lambda_4 \}
	\end{align}
\end{linenomath}
where $\lambda_i$ are the square roots of the eigenvalues of the matrix $\rho (\sigma_{y} \otimes \sigma_{y}) \rho^{*} (\sigma_{y} \otimes \sigma_{y})$, in descending order and $\rho^*$ is the elementwise complex conjugate of the density matrix $\rho$.

\section{Additional quantum state transfer and remote entanglement measurements}
\begin{figure}
	\centering	
	\includegraphics[width=\textwidth]{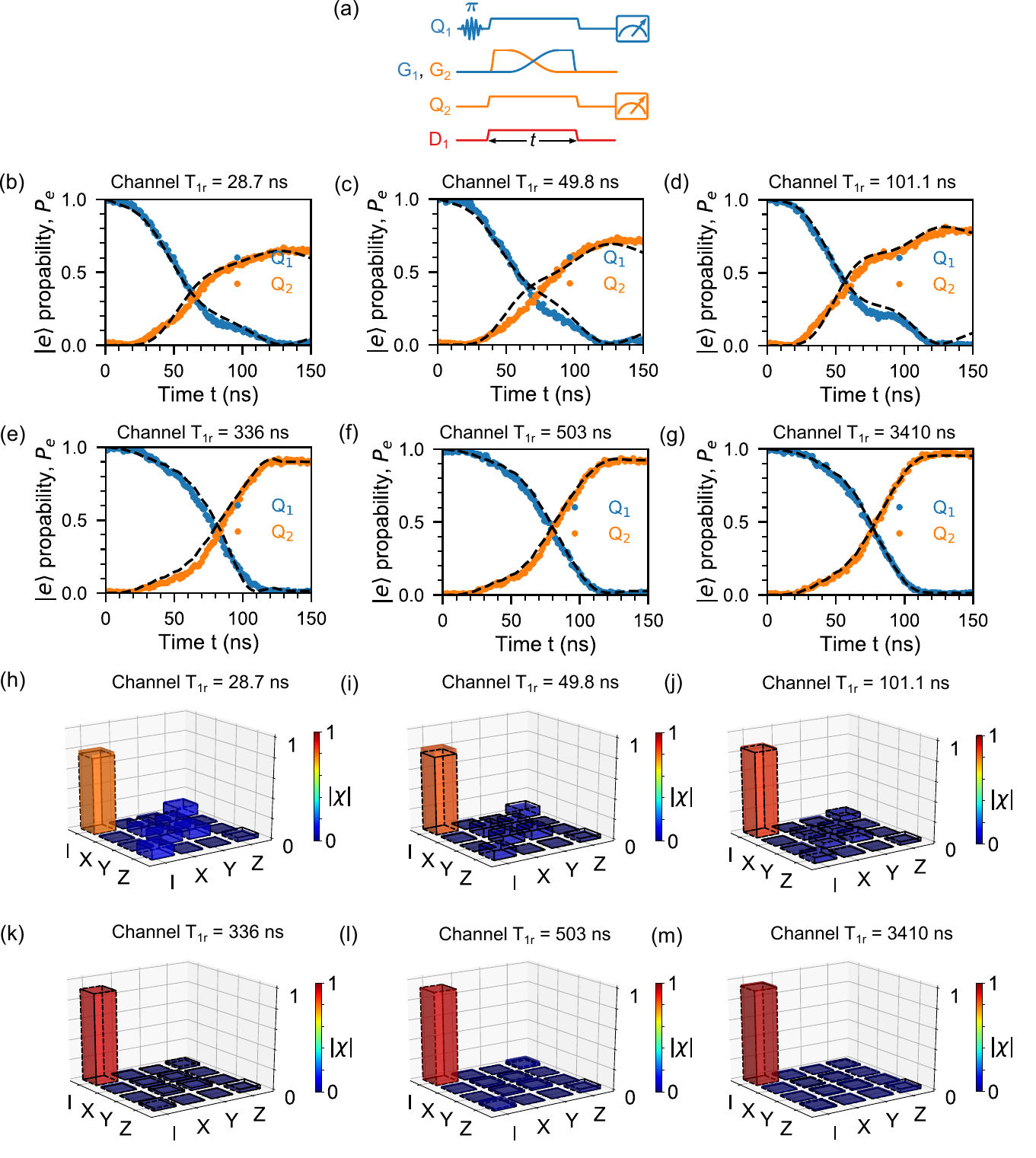}
		\caption{
			\label{fig:s_adb_QST}
			Quantum state transfer using the adiabatic protocol.
			(a) Control pulse sequence.
			(b-g) Adiabatic state transfer between qubits $Q_1$ and $Q_2$, measured with different dissipation settings for the resonant channel mode, quantified by the resonant mode lifetime $T_{1r}$. Blue (orange) circles represent simultaneously measured excited state populations of $Q_1$ ($Q_2$) at time $t$.
			(h-m) Quantum process tomography at the maximum transfer efficiency point for each dissipation setting in panels b-g. In all panels, dashed lines are the results from master equation simulations, accounting for channel dissipation and qubit imperfections.
			}
	\centering
\end{figure}

\begin{figure}
	\centering		
	\includegraphics[width=\textwidth]{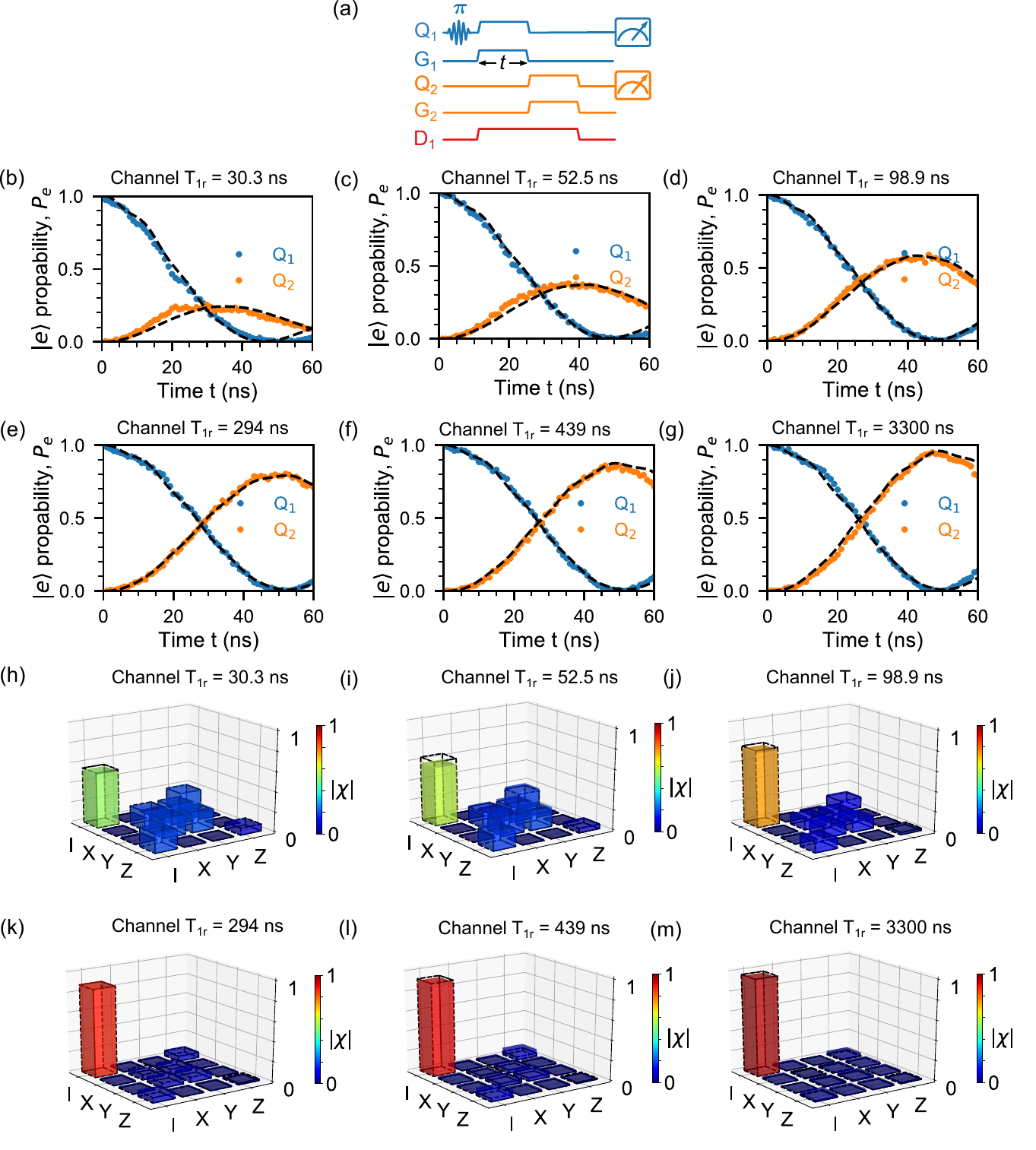}
	\caption{
		\label{fig:s_relay_QST}
		Quantum state transfer using the relay method.
		(a) Control pulse sequence.
		(b-g) Quantum state transfer from $Q_1$ to $Q_2$ using the resonant channel mode as a relay, measured with different dissipation settings for the resonant channel mode, quantified by the resonant mode lifetime $T_{1r}$. Blue (orange) circles represent simultaneously  measured excited state populations of $Q_1$ ($Q_2$) versus swap time $t$.
		(h-m) Quantum process tomography at the maximum transfer efficiency point for each dissipation setting in panels b-g. In all panels, dashed lines are the results from master equation simulations, accounting for channel dissipation and qubit imperfections.
	}
	\centering
\end{figure}

\begin{figure}
	\centering		
	\includegraphics[width=\textwidth]{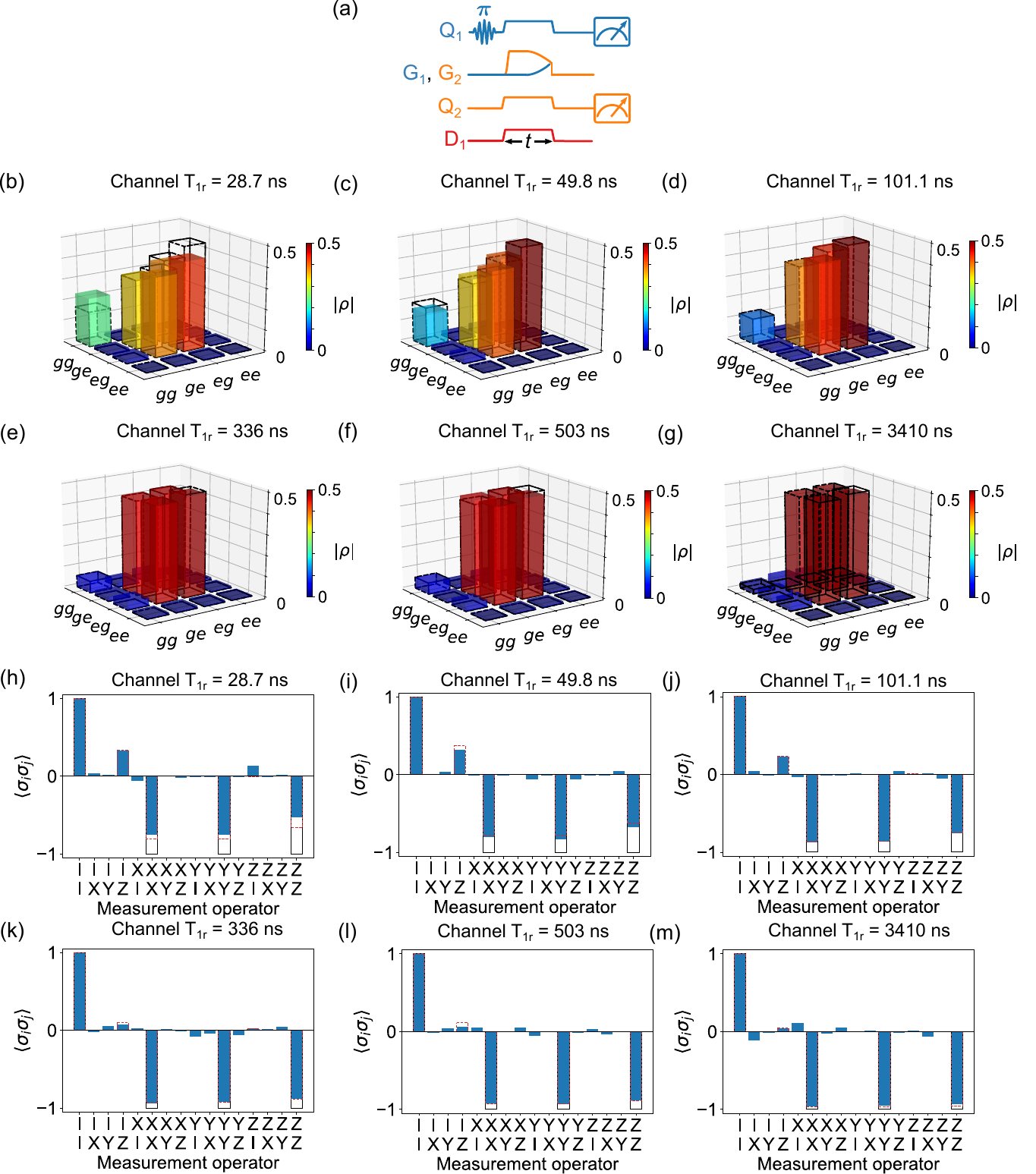}
	\caption{
		\label{fig:s_adb_Bell}
		Remote entanglement using the adiabatic protocol.
		(a) Control pulse sequence.
		(b-g) Reconstructed density matrix of the Bell states generated using the adiabatic protocol, measured with different dissipation settings for the resonant channel mode, quantified by the resonant mode lifetime $T_{1r}$.
		(h-m) Expectation values for the two-qubit Pauli operators $\langle \sigma_i \sigma_j \rangle$ for the Bell state density matrix in panels b-g. Solid lines show the expectation values for the ideal Bell singlet state $| \psi^{-} \rangle = \left (|e0g\rangle - |g0e\rangle\right )/\sqrt{2}$. In all panels, dashed lines are the results from master equation simulations, accounting for channel dissipation and qubit imperfections.
	}
	\centering
\end{figure}

\begin{figure}
	\centering
	\includegraphics[width=\textwidth]{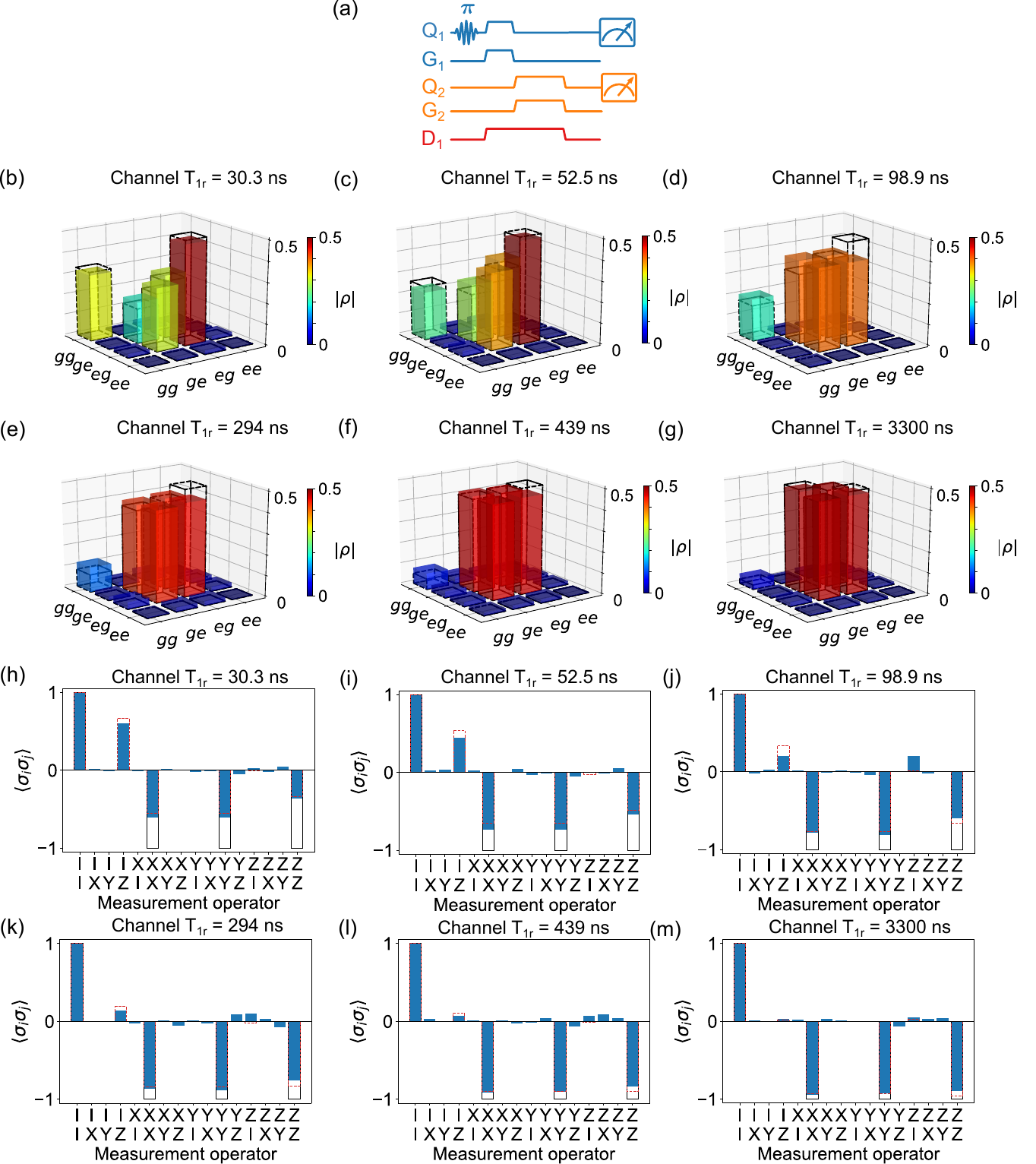}
	\caption{
		\label{fig:s_relay_Bell}
		Remote entanglement using the relay method.
		(a) Control pulse sequence.
		(b-g) Reconstructed density matrix of the Bell states generated with the relay method, measured with different dissipation settings for the resonant channel mode, quantified by the resonant mode lifetime $T_{1r}$.
		(h-m) Expectation values for the two-qubit Pauli operators $\langle \sigma_i \sigma_j \rangle$ for the Bell state density matrix in panels b-g. Solid lines show the expectation values for the ideal Bell singlet state $| \psi^{-} \rangle = \left (|e0g\rangle - |g0e\rangle\right )/\sqrt{2}$. In all panels, dashed lines are the results from master equation simulations, accounting for channel dissipation and qubit imperfections.
	}
	\centering

\end{figure}

In Fig.~\ref{fig:s_adb_QST}--\ref{fig:s_relay_Bell}, we show additional measurements similar to those shown in Fig.~3a,b of the main text, for other dissipation settings in the channel mode. These measurements were made using both the adiabatic protocol and the relay method. Results from a master equation simulation, accounting for channel dissipation as well as qubit imperfections are shown as well.

%%%%% OTHER SUPPLEMENTARY FIGURES
\clearpage
\renewcommand{\baselinestretch}{1.}

\begin{table}[H]
	\begin{center}
		\begin{tabular}{p{9cm}ccc}
			\hline
			\hline
			Qubit parameters & \ & Qubit 1 & Qubit 2 \\
			\hline
			Qubit maximum frequency, $\omega_{ge}^{max}/2\pi$ (GHz) & \ &  6.239 & 6.132 \\
			Qubit idle frequency, $\omega_{ge}^{idle}/2\pi$ (GHz) & \ & 5.504 & 5.419 \\
			Qubit capacitance, $C_q$ (design value) (fF) & \ & 90 & 90 \\
			Qubit SQUID inductance, $L_q$ (nH) & \ & 7.2 & 7.5 \\
			Qubit anharmonicity, $\alpha/2\pi$ (MHz)  & \ & -168 & -171 \\
			Qubit intrinsic lifetime, $T_{1,\mathrm{int}}$ ($\mu$s) & \ & 11.5(5) & 9.1(2) \\
			Qubit Ramsey dephasing time, $T_{2,\mathrm{Ramsey}}$ ($\mu$s) & \ & 1.11(3) & 1.15(3) \\
			Qubit spin-echo dephasing time, $T_{2E}$ ($\mu$s) & \ & 4.09(5) & 3.54(4) \\
			$|g\rangle$ state readout fidelity, $F_g$ & \ & 0.966(3) & 0.959(4) \\
			$|e\rangle$ state readout fidelity, $F_e$ & \ & 0.881(5) & 0.888(8) \\
			Readout resonator frequency, $\omega_{r}/2\pi$ (GHz) & \ & 6.361 & 6.415 \\
			Readout resonator quality factor, $Q_r$ & \ & $6.9\times10^3$ & $6.4\times10^3$ \\
			Readout dispersive shift, $\chi_r/2\pi$ (MHz) & \ & 0.15 & 0.15 \\
			\hline
			\hline
			\  & \  & \  & \ \\
			\hline
			\hline
			Flux-tunable couplers parameters & \ & Coupler $G_1$ & Coupler $G_2$ \\
			\hline
			Coupler junction inductance, $L_{T}$ (nH) & \ & 0.61 & 0.61 \\
			Coupler grounding inductance, $L_{g}$ (design value) (nH) & \ & 0.2 & 0.2 \\
			\hline
			\hline
			\  & \  & \  & \ \\
			\hline
			\hline
			Tunable switch parameters & \ & Coupler $D_1$ & Coupler $D_2$ \\
			\hline
			Coupler SQUID inductance, $L_{J}$ (nH) & \ & 0.34 & 0.34 \\
			Coupler SQUID capacitance, $C_{J}$ (fF) & \ & 125 & 125 \\
			Coupler grounding capacitance, $C_{g}$ (design value) (fF) & \ & 100 & 100\\
			\hline
			\hline
		\end{tabular}
	\end{center}
	\caption{\label{table:parameters}
		Device parameters for the two qubits, the flux-tunable couplers connecting each qubit to the channel, and the DC SQUID tunable couplers making up the tunable switch that couple the channel to an external $50~\Omega$ load.}
\end{table}

\renewcommand{\baselinestretch}{1.5}

\clearpage
\bibliography{bibliography}